\documentclass[a4paper,14pt]{article}

\usepackage[margin=1.5in]{geometry}

\usepackage[english]{babel}
\usepackage[T1]{fontenc}
\usepackage[utf8]{inputenc}
\usepackage{lmodern}
\usepackage{microtype}

\usepackage{bbm}
\usepackage{amsmath}
\usepackage{amssymb}
\usepackage{enumerate}
\usepackage{amsthm}

\usepackage{xcolor}

\usepackage{listings}

\usepackage{graphicx}
\usepackage{epsfig}

\usepackage{url}
\usepackage[backend=bibtex,citestyle=numeric]{biblatex}
\usepackage{hyperref}

\usepackage{comment} 

\title{The DTM-signature for a geometric comparison of metric-measure spaces from samples\footnote{This work was partially supported by the ANR project TopData and GUDHI}}

\author{Claire Br\'echeteau\\\small{Universit\'e Paris-Sud -- Inria Saclay // Universit\'e Paris-Saclay, France}\\
  \texttt{claire.brecheteau@inria.fr}}

\newtheoremstyle{break}  
  {\topsep}   
  {\topsep}   
  {\itshape}  
  {0pt}       
  {\bfseries} 
  {.}         
  {\newline}  
  {}          

\newtheoremstyle{break2}  
  {\topsep}   
  {\topsep}   
  {\itshape}  
  {0pt}       
  {\bfseries} 
  {}         
  {\newline}  
  {}          

\theoremstyle{break}
\newtheorem{thmm}{Theorem}
\newtheorem{dff}[thmm]{Definition}
\newtheorem{propp}[thmm]{Proposition}
\newtheorem{corr}[thmm]{Corollary}
\newtheorem{lmm}[thmm]{Lemma}

\theoremstyle{break2}
\newtheorem{Preuve}{Proof}

\theoremstyle{plain}
\newtheorem{Remarque}{Remark}
\newtheorem{exx}[thmm]{Example}

\newenvironment{pvv}{\begin{Preuve}\rm}%
  {$\blacksquare$\end{Preuve}}
\newenvironment{rqq}{\begin{Remarque}\rm}%
  {\end{Remarque}}

\newcommand{\thm}{\begin{thmm}}
\newcommand{\ethm}{\end{thmm}}
\newcommand{\lm}{\begin{lmm}}
\newcommand{\elm}{\end{lmm}}
\newcommand{\ex}{\begin{exx}}
\newcommand{\eex}{\end{exx}}
\newcommand{\df}{\begin{dff}}
\newcommand{\edf}{\end{dff}}
\newcommand{\prop}{\begin{propp}}
\newcommand{\eprop}{\end{propp}}
\newcommand{\pv}{\begin{pvv}}
\newcommand{\epv}{\end{pvv}}
\newcommand{\rem}{\begin{rqq}}
\newcommand{\erem}{\end{rqq}}
\newcommand{\cor}{\begin{corr}}
\newcommand{\ecor}{\end{corr}}

\usepackage{bbm}
\usepackage{color, colortbl}

\lstdefinestyle{cust}{
language=python,
commentstyle=\ttfamily,
basicstyle=,
escapeinside={\%*}{*)},
frame=single,
keepspaces=true,
keywordstyle=\bfseries,
morekeywords={*,Input,Output},
}
\usepackage{xcolor}
\newcommand{\R}{\mathbbm R}
\newcommand{\N}{\mathbbm N}
\newcommand{\E}{\mathbbm E}
\newcommand{\G}{\mathbbm G}
\newcommand{\1}{\mathbbm 1}
\newcommand{\dd}{\rm d\it}
\newcommand{\x}{\mathcal X}
\newcommand{\y}{\mathcal Y}
\newcommand{\z}{\mathcal Z}
\newcommand{\MM}{\mathcal M}
\newcommand{\LLL}{\mathcal L}
\newcommand{\Diam}{\mathcal D}
\newcommand{\T}{\mathcal T}
\newcommand{\SSS}{\mathcal S}
\newcommand{\B}{\rm B\it}
\newcommand{\Leb}{\rm Leb\it}
\newcommand{\Reach}{\rm Reach\it}
\newcommand{\Supp}{\rm Supp \it}
\newcommand{\qu}{\rm q\it}
\newcommand{\Sk}{\rm Sk\it}
\newcommand{\p}{\mathbbm P}

\definecolor{Gray}{gray}{0.9}

\bibliography{biblio}

\begin{document}

\maketitle

\begin{abstract}
In this paper, we introduce the notion of DTM-signature, a measure on $\R_+$ that can be associated to any metric-measure space. This signature is based on the distance to a measure (DTM) introduced by Chazal, Cohen-Steiner and M\'erigot. It leads to a pseudo-metric between metric-measure spaces, upper-bounded by the Gromov-Wasserstein distance. Under some geometric assumptions, we derive lower bounds for this pseudo-metric.

Given two $N$-samples, we also build an asymptotic statistical test based on the DTM-signature, to reject the hypothesis of equality of the two underlying metric-measure spaces, up to a measure-preserving isometry. We give strong theoretical justifications for this test and propose an algorithm for its implementation.
\end{abstract}

\section{Introduction}
Among the variety of data available, from astrophysics to biology, including social networks and so on, many come as sets of points from a metric space. A natural question, given two sets of such data is to decide whether they are similar, that is whether they come from the same distribution, whether their shape are close, or not. This comparison may be compromised when the data are not embedded into the same space, or if the two systems of coordinates in which the data are represented are different. To overcome this issue, a natural idea is to forget about this embedding and only consider the set of points together with the distances between pairs.  A natural framework to compare data is then to assume that they come from a measure on a metric space and to consider two such metric-measure spaces as being the same when they are equal up to some isomorphism, as defined below.

\df[mm-space]
A \textbf{metric-measure space} (\textbf{mm-space}) is a triple $(\x,\delta,\mu)$, with $\x$ a set, $\delta$ a metric on $\x$ and $\mu$ a probability measure on $\x$ equipped with its Borel $\sigma$-algebra. 
\edf

\df[Isomorphism between mm-spaces]
Two mm-spaces $(\x,\delta,\mu)$ and $(\y,\gamma,\nu)$ are said to be \textbf{isomorphic} if their exist Borel sets $\x_0\subset\x$ and $\y_0\subset\y$ such that $\mu(\x\backslash\x_0)=0$ and $\nu(\y\backslash\y_0)=0$, and some one-to-one and onto isometry $\phi:\x_0\rightarrow \y_0$ preserving measures, that is satisfying $\nu(\phi(A\cap\x_0))=\mu(A\cap\x_0)$ for any Borel set $A$ of $\x$.
Such a map $\phi$ is called an \textbf{isomorphism} between the mm-spaces $(\x,\delta,\mu)$ and $(\y,\gamma,\nu)$.
\edf

In this paper, we first address the question of the comparison of general mm-spaces, up to an isomorphism. In other terms, we aim at designing a metric or at least a pseudo-metric on the quotient space of mm-spaces by the relation of isomorphism. A suitable pseudo-distance should be stable under some perturbations, under sampling, discriminative and easy to implement when dealing with discrete spaces.

A first characterisation of mm-spaces is given in \cite{Gromov}. In its Theorem $3\frac{1}{2}.5$, Gromov proves that any mm-space can be recovered, up to an isomorphism, from the knowledge, for all size $N$, of the distribution of the $N\times N$-matrix of distances associated to a $N$-sample. More recently, in \cite{Memoli}, Mémoli proposes metrics on the quotient space of mm-spaces by the relation of isomorphism, the Gromov--Wasserstein distances.

\df[Gromov--Wasserstein distance]
The \textbf{Gromov--Wasserstein distance} between two mm-spaces $(\x,\delta,\mu)$ and $(\y,\gamma,\nu)$ with parameter $p\in[1,\infty)$ denoted $GW_p(\x,\y)$ is defined by the expression:
\[\inf_{\pi\in\Pi(\mu,\nu)}\frac{1}{2}\left(\int_{\x\times\y}\int_{\x\times\y}\left(\Gamma_{\x,\y}\left(x,y,x',y'\right)\right)^p\pi(\dd x\times \dd y)\pi(\dd x'\times \dd y')\right)^{\frac{1}{p}},\]
with $\Gamma_{\x,\y}(x,y,x',y')=\left|\delta(x,x')-\gamma(y,y')\right|$.
Here $\Pi(\mu,\nu)$ stands for the set of \textbf{transport plans} between $\mu$ and $\nu$, that is the set of Borel probability measures $\pi$ on $\x\times\y$ satisfying 
$\pi(A\times\y)=\mu(A)$ and $\pi(\x\times B)=\nu(B)$ for all Borel sets $A$ in $\x$ and $B$ in $\y$.
\edf

Unfortunately, even when dealing with discrete mm-spaces, the computation of these Gromov--Wasserstein distances is extremely costly. An alternative is to build a \textbf{signature} from each mm-space, that is an object invariant under isomorphism. The mm-spaces are then compared through their signatures. In \cite{Memoli}, M\'emoli gives an overview of such signatures, as for instance shape distribution, eccentricity or what he calls local distribution of distances.

In this paper, we introduce a new signature that is a probability measure on $\R$, and we propose to compare such signatures using Wasserstein distances \cite{Villani}.

\df[Wasserstein distance]
\label{def Wasserstein}
The \textbf{Wasserstein distance} of parameter $p\in[1,\infty)$ between two Borel probability measures $\mu$ and $\nu$ over the same metric space $(\x,\delta)$ is defined as:
\[W_p(\mu,\nu)=\inf_{\pi\in\Pi(\mu,\nu)}\left(\int_{\R^2}\delta^p(x,y)\dd\pi(x,y)\right)^{\frac{1}{p}}.\]
\edf
For two probability measures $\mu$ and $\nu$ over $\R_+$, the $L_1$-Wasserstein distance can be rewritten as the $L_1$-norm between the cumulative distribution functions of the measures, $F_{\mu}:t\mapsto\mu((-\infty,t])$ and $F_{\nu}$, or as well, as the $L_1$-norm between the quantile functions, $F_\mu^{-1}:s\mapsto\inf\{x\in\R\,|\,F(x)\geq s\}$ and $F_\nu^{-1}$. Thus, the computation of the $L_1$-Wasserstein distance between empirical measures is easy, in $O(N\log(N))$ for two empirical measures from subsets of $\R$ of size $N$, the complexity of a sort.

Shape signatures are widely used for classification or pre-classification tasks; see for instance \cite{Osada}. With a more topological point of view, persistence diagrams have been used for this purpose in \cite{ChazalMemoli, Oudot}. But, as far as we know, the construction of well-founded statistical tests from signatures to compare mm-spaces has not been considered among the literature. This is the second problem focussed in this paper.

Recall that a \textbf{statistical test} is a random variable $\phi_N$ taking values in $\{0,1\}$. More precisely $\phi_N$ is a function of $N$ random data from a distribution $\LLL_\theta$ depending on some unknown parameter $\theta$ in some set $\Theta$. It is associated to two hypotheses $H_0$``$\theta\in\Theta_0$'' and $H_1$``$\theta\in\Theta_1$'' with $\Theta_0$ and $\Theta_1$ disjoint subsets of $\Theta$. Ideally, we would like the test $\phi_N$ to be equal to 1 if $\theta$ is in $\Theta_1$ and to be 0 if $\theta$ is in $\Theta_0$.

The quality of a statistical test is measured in terms of its \textbf{type I error}, that is the function defined for all $\theta_0$ in $\Theta_0$ by $\p_{\theta_0}(\phi_N=1)$, the probability of pretending $\theta$ to be in $\Theta_1$ when $\theta=\theta_0$ is actually in $\Theta_0$. Moreover, a test is of \textbf{level} $\alpha\in(0,1)$ if its type I error is upper-bounded by $\alpha$, that is $\p_{\theta_0}(\phi_N=1)\leq\alpha$ for all $\theta_0$ in $\Theta_0$.
Two statistical tests with a fixed level $\alpha\in(0,1)$ can be compared through their \textbf{type II error}, that is the function defined for all $\theta_1$ in $\Theta_1$ by $\p_{\theta_1}(\phi_N=0)$, the probability of pretending $\theta$ to be in $\Theta_0$ when $\theta=\theta_1$ is actually in $\Theta_1$. See \cite{Bickel} for a reference on statistical tests.

In this article, we build a \textbf{test of asymptotic level} $\alpha$, that is a test $\phi_N$ such that for any $\theta_0$ in $\Theta_0$, $\p_{\theta_0}(\phi_N=1)\rightarrow\alpha$ when the size of the sample $N$ goes to $\infty$. Moreover, the set $\Theta$ we consider is the set of couples of mm-spaces $((\x,\delta,\mu),(\y,\gamma,\nu))$. The set $\Theta_0$ is the subset of $\Theta$ made of couples of two isomorphic spaces: $\Theta_0=\{((\x,\delta,\mu),(\y,\gamma,\nu))\in\Theta\,|\,(\x,\delta,\mu)\text{ and }(\y,\gamma,\nu)\text{ are isomorphic}\}$, and $\Theta_1=\Theta\backslash\Theta_0$.

Such a test generalises two-sample tests, from the precursor Kolmogorov-Smirnov test to the more recent tests in \cite{Gretton} or \cite{Cazals}. Our test does not depend on the embedding of the data and keeps a track of the geometry in some way, a point of view that has already been taken in the context of density estimation \cite{Luxburg}. Thus, it could be of interest for proteins, 3D-shape comparison, etc.\\

Concretely, in this paper, we propose a new signature based on the distance to a measure (DTM) introduced in \cite{Merigot}, the \textbf{DTM-signature}. This signature is invariant under isomorphism and easy to compute. We prove its stability with respect to the so-called Gromov-Wasserstein and Wasserstein distances with parameter $p=1$. It leads to a stability under sampling, at least for the Euclidean space $\R^d$. After deriving frameworks under which the knowledge of the distance to a measure determines the measure, we prove discriminative properties for the DTM-signature by deriving lower bounds for the $L_1$-Wasserstein distance between two such signatures, under various assumptions. Finally, from two $N$-samples, we derive a statistical test, based on bootstrap methods, to reject or not the hypothesis of equality of the two underlying metric-measure spaces, up to a measure-preserving isometry. This test comes with an easy-to-implement algorithm, and a strong theoretical justification.

The DTM-signature depends on some parameter $m\in(0,1)$. It thus offers a variety of new fictures, as well as new lower-bounds for the Gromov-Wasserstein distance. As for the statistical test, it presents the advantage of not depending on the embedding of the data, only the knowledge of the distances between points is required. In this sense, it is new. The justification of the valitidy of the test with the use of the Wasserstein distance is quite new as well, and still poorly used; see \cite{Helene} for another use.\\

The paper is organized as follows. Section \ref{The distance to a measure to dicriminate between measures} is devoted to the distance to a measure. An accent is put on its discriminative properties. The DTM-signature is then introduced in Section \ref{The DTM-signature to discriminate between metric-measure spaces}. The question of discrimination of two mm-spaces is also discussed. For this purpose, we derive lower bounds for our pseudo-distance, the $L_1$-Wasserstein distance between the two DTM-signatures. Finally, in Section \ref{An algorithm to compare metric-measure spaces from samples} we introduce the test of isomorphism, propose an algorithm for its implementation and then give some theoretical results to ensure the validity of the procedure. Numerical illustrations are given in Section \ref{illustration}.

\section{The distance to a measure to discriminate between measures}
\label{The distance to a measure to dicriminate between measures}
Let $(\x,\delta)$ be a metric space, equipped with a Borel probability measure $\mu$. Given $m$ in $[0,1]$, the \textbf{pseudo-distance function} is defined at any point $x$ of $\x$, by:
\[\delta_{\mu,m}(x)=\inf\{r>0\,|\,\mu(\overline{\B}(x,r))>m\}.\]

The function \textbf{distance to the measure} $\mu$ with mass parameter $m$ and denoted $\dd_{\mu,m}$ is then defined for all $x$ in $\x$ by:
\[\dd_{\mu,m}(x)=\frac{1}{m}\int_{l=0}^m\delta_{\mu,l}(x)\,\dd l.\]

The distance to a measure is a generalisation of the function distance to a compact set; see \cite{Merigot}.
This function is continuous with respect to the mass parameter $m$, and Lipschitz with respect to $\mu$.

\prop[Stability, in \cite{Merigot} for $\R^d$, in \cite{Buchet} for metric spaces]
\label{Stab DTM}
For two mm-spaces $(\x,\delta,\mu)$ and $(\y,\delta,\nu)$ embedded into the same metric space, we have that
\[\|\dd_{\mu,m}-\dd_{\nu,m}\|_{\infty,\x\cup\y}\leq\frac1m W_1(\mu,\nu).\]
\eprop

Moreover, for some empirical measure $\hat\mu_N=\frac1N\sum_{i=1}^N\delta_{X_i}$ on a metric space $(\x,\delta)$, the distance to the measure $\hat\mu_N$ with mass parameter $\frac{k}{N}$ for some $k$ in $[\![0,N]\!]$ at a point $x$ of $\x$ satisfies:
 \[\dd_{\hat\mu_N,\frac{k}{N}}(x)=\frac{1}{k}\sum_{i=1}^k\delta(X^{(i)},x),\]
where $X^{(1)}$, $X^{(2)}$, \ldots $X^{(k)}$ are $k$ nearest neighbours of $x$ among the $N$ points $X_1$, $X_2$, \ldots $X_N$.

The distance to the measure $\hat\mu_N$ is thus equal to the mean of the distances to $k$-nearest neighbours. In particular, in this case, the computation of the DTM boils down to the computation of the first $k$-nearest neighbours.\\

The question of determining if the knowledge of the distance to a measure leads to the knowledge of the measure itself is a natural question. Some work has been done in this direction for discrete measures; see \cite{Buchet}. In the following, we propose results in different settings.

\prop
Let $(\x,\delta)$ be a metric space, and $\MM_1(\x)$ be the set of Borel probability measures over $(\x,\delta)$. We define the maps $\phi$ and $\psi$ for all $\mu$ in $\MM_1(\x)$ by:
\[\phi(\mu)=\left(\dd_{\mu,m}(x)\right)_{m\in[0,1],\,x\in\x}\]
and
\[\psi(\mu)=\left(\mu\left(\overline{\B}(x,r)\right)\right)_{r\in\R_+,\,x\in\x}.\]
Then, the map $\phi$ is injective if and only if the map $\psi$ is injective.
\eprop
\pv
From the definition of $\delta_{\mu,m}(x)$, we have:
\[\mu\left(\overline{\B}(x,r)\right)=\inf\{m\geq0\,|\,\delta_{\mu,m}(x)>r\}.\]
Moreover, since $m\rightarrow\delta_{\mu,m}(x)$ is right-continuous, after the differentiation the distance-to-a-measure function with respect to $m$, we have:
\[m\frac{\partial}{\partial m}\dd_{\mu,m}(x)+\dd_{\mu,m}(x)=\delta_{\mu,m}(x).\]\epv

It means that in spaces on which measures are determined by their values on balls, the measures are determined by the knowledge of the distance-to-a-measure functions for all parameters $m$ in $[0,1]$, on all $x$ in $\x$. Remark that the Euclidean space $\R^d$ satisfies such a condition, but this is not the case of every metric space, as explained in \cite{Buet}.\\
\label{Uniform measures on non-empty open subsets}

Under the following specific framework, we will establish a stronger identifiability result.

For $O$ a non-empty bounded open subset of $\R^d$, we define the \textbf{uniform measure} $\mu_O$ for all Borel set $A$ of $\R^d$, by:
\[\mu_O(A)=\frac{\Leb_d(O\cap A)}{\Leb_d(O)},\]
with $\Leb_d$ the Lebesgue measure on $\R^d$.

We also define the \textbf{medial axis} of $O$, $\MM(O)$ as the set of points in $O$ having at least two projections onto $\partial O$. That is,
\[\MM(O)=\{y \in O\,|\,\exists\, x',x''\in\partial O,\, x'\neq x'',\, \|y-x'\|_2=\|y-x''\|_2=\dd(y,\partial O)\},\] with $\dd(y,\partial O)=\inf\{\|x-y\|_2\,|\,x\in\partial O\}$.

Its \textbf{reach}, $\Reach(O)$, is the distance between its boundary $\partial O$ and its medial axis $\MM(O)$. That is,
\[\Reach(O)=\inf\{\|x-y\|_2\,|\,x\in\partial O,\, y\in\MM(O)\}.\]

If $K$ is a compact subset of $\R^d$, it is standard to define its reach as $\Reach\left(K^c\right)$, the reach of its complement in $\R^d$. See \cite{Federer} to get more familiar with these notions.

\prop
\label{inject measure}
Let $O$ and $O'$ be two non-empty bounded open subsets of $\R^d$ with positive reach, such that $O=\left(\overline{O}\right)^{\circ}$ and $O'=\left({\overline{O'}}\right)^{\circ}$. Let $m$ be some positive constant satisfying \[m\leq\min\left(\Reach(O)^d,\Reach(O')^d\right)\frac{\omega_d}{\Leb_d(O)},\]
with $\omega_d=\Leb_d(\B(0,1))$, the Lebesgue volume of the unit $d$-dimensional ball.
If for all $x$ in $\R^d$ 
\[\dd_{\mu_O,m}(x)=\dd_{\mu_{O'},m}(x),\]
then $\mu_O=\mu_{O'}$.
\eprop
\pv This is a straightforward consequence of Proposition \ref{id of measures}, in the Appendix. The proof relies on the fact that the set of points in $\R^d$ minimizing the distance to the measure $\mu_O$ is equal to $\{x\in O\,|\,\dd_{\partial O}(x)\geq\epsilon(m,O)\}$ with $\epsilon(m,O)=\left(\frac{m\Leb_d(O)}{\omega_d}\right)^\frac1d$, providing that the set is non-empty. Then, if $\Reach(O)$ is not smaller than $\epsilon(m,O)$, $O$ equals to the set of points at distance smaller than $\epsilon(m,O)$ from $\{x\in O\,|\,\dd_{\partial O}(x)\geq\epsilon(m,O)\}$. Thus, the measure $\mu_O$ can be recovered. We use the notion of \textbf{skeleton} in \cite{Lieutier} for some details in the proof.
\epv
It means that for $m$ small enough, the knowledge of the distance to a measure at any point $x$ in $\R^d$ for two measures $\mu_O$ and $\mu_{O'}$ is discriminative.

\section{The DTM-signature to discriminate between metric-measure spaces}
\label{The DTM-signature to discriminate between metric-measure spaces}
From the distance-to-a-measure function, we derive a new signature.

\df[DTM-signature]
The \textbf{DTM-signature} associated to some mm-space $(\x,\delta,\mu)$, denoted $\dd_{\mu,m}(\mu)$, is the distribution of the real-valued random variable $\dd_{\mu,m}(X)$ where $X$ is some random variable of law $\mu$.
\edf

The DTM-signature turns out to be stable in the following sense.

\prop
\label{Stab wr GW}
We have that:
\[W_1(\dd_{\mu,m}(\mu),\dd_{\nu,m}(\nu))\leq \frac1m GW_1(\x,\y).\]
\eprop

\pv Proof in the Appendix, in Section \ref{section stab}. The proof is relatively similar to the ones given by M{\'e}moli in \cite{Memoli} for other signatures.
\epv

It follows directly that two isomorphic mm-spaces have the same DTM-signature.
Whenever the two mm-spaces are embedded into the same metric space, we also get stability with respect to the $L_1$-Wasserstein distance.

\prop
\label{Stab wr W}
If $(\x,\delta,\mu)$ and $(\y,\delta,\nu)$ are two metric spaces embedded into some metric space $(\z,\delta)$, then we can upper bound $W_1(\dd_{\mu,m}(\mu),\dd_{\nu,m}(\nu))$ by
\[W_1(\mu,\nu)+\min\left\{\|\dd_{\mu,m}-\dd_{\nu,m}\|_{\infty,\Supp(\mu)},\|\dd_{\mu,m}-\dd_{\nu,m}\|_{\infty,\Supp(\nu)}\right\},\]
and more generally by
\[\left(1+\frac1m\right) W_1(\mu,\nu).\]
\eprop

\pv
First remark that:
\begin{align*}
W_1(\dd_{\mu,m}(\mu),\dd_{\nu,m}(\mu))&\leq\int_{\x}\left|\dd_{\mu,m}(x)-\dd_{\nu,m}(x)\right|\dd\mu(x)\\
&\leq\|\dd_{\mu,m}-\dd_{\nu,m}\|_{\infty,\Supp(\mu)}.
\end{align*}
Then, for all $\pi$ in $\Pi(\mu,\nu)$:
\[W_1(\dd_{\nu,m}(\mu),\dd_{\nu,m}(\nu))\leq\int_{\x\times\y}\left|\dd_{\nu,m}(x)-\dd_{\nu,m}(y)\right|\dd\pi(x,y).\]
Thus, since $\dd_{\nu,m}$ is 1-Lipschitz:
\[W_1(\dd_{\nu,m}(\mu),\dd_{\nu,m}(\nu))\leq W_1(\mu,\nu).\]
We use Proposition \ref{Stab DTM} to conclude.
\epv

The DTM-signature is stable but unfortunately does not always discriminates between mm-spaces. Indeed, in the following counter-example from \cite{Memoli} (example 5.6), there are two non-isomorphic mm-spaces sharing the same signatures for all values of $m$.

\ex
\label{deux graphes}
We consider two graphs made of 9 vertices each, clustered in three groups of 3 vertices, such that each vertex is at distance 1 exactly to each vertex of its group and at distance 2 to any other vertex. We assign a mass to each vertex, the distribution is the following, for the first graph: 
\[\mu=\left\{\left(\frac{23}{140},\frac{1}{105},\frac{67}{420}\right),\left(\frac{3}{28},\frac{1}{28},\frac{4}{21}\right),\left(\frac{2}{15},\frac{1}{15},\frac{2}{15}\right)\right\},\]
and for the second graph:
\[\nu=\left\{\left(\frac{3}{28},\frac{1}{15},\frac{67}{420}\right),\left(\frac{2}{15},\frac{4}{21},\frac{1}{105}\right),\left(\frac{23}{140},\frac{2}{15},\frac{1}{28}\right)\right\}.\]
The mm-spaces ensuing are not isomorphic since any one-to-one and onto measure-preserving map would send at least one couple of vertices at distance 1 to each other, to a couple of vertices at distance 2 to each other, thus it would not be an isometry.

Moreover, remark that the DTM-signatures associated to the graphs are equal since the total mass of each cluster is exactly equal to $\frac13$.

\begin{figure}[h!]
 \begin{minipage}[h]{.46\linewidth}
  \centering\includegraphics[scale=1.2]{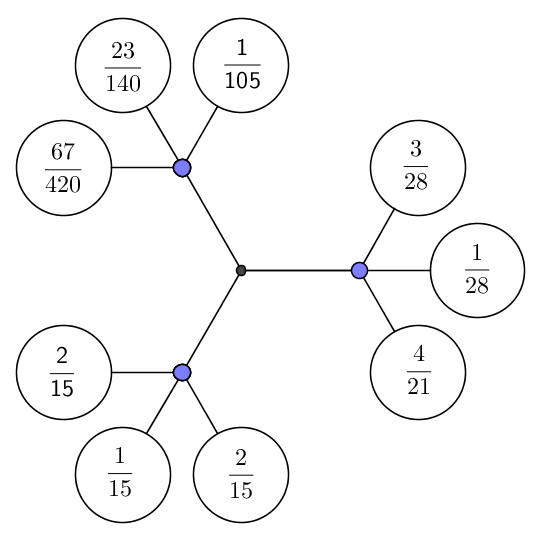}
  \caption{$\mu$}
 \end{minipage} \hfill
 \begin{minipage}[h]{.46\linewidth}
  \centering\includegraphics[scale=1.2]{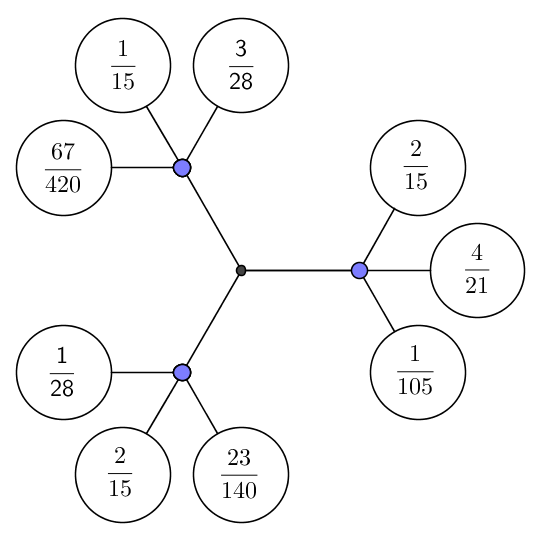}
  \caption{$\nu$}
 \end{minipage}
\end{figure}
\eex

Nevertheless, the signature can be discriminative in some cases. In the following, we give lower bounds for the $L_1$-Wasserstein distance between two signatures under three different alternatives.

\subsection{When the distances are multiplied by some positive real number $\lambda$}
Let $\lambda$ be some positive real number. The DTM-signature discriminates between two mm-spaces isomorphic up to a dilatation of parameter $\lambda$, for $\lambda\neq1$.

\prop
Let $(\x,\delta,\mu)$ and $(\y,\gamma,\nu)=(\x,\lambda\delta,\mu)$ be two mm-spaces. We have
\[W_1(\dd_{\mu,m}(\mu),\dd_{\nu,m}(\nu))=|1-\lambda|\,\E_{\mu}[\dd_{\mu,m}(X)],\]
for $X$ a random variable of law $\mu$.
\eprop
\pv
First remark that $F^{-1}_{\dd_{\nu,m}(\nu)}=\lambda F^{-1}_{\dd_{\mu,m}(\mu)}$.
Then,
\begin{align*}
W_1(\dd_{\mu,m}(\mu),\dd_{\nu,m}(\nu)) &= \int_0^1\left|F^{-1}_{\dd_{\mu,m}(\mu)}(s)-F^{-1}_{\dd_{\nu,m}(\nu)}(s)\right|\,\dd s\\
&=|1-\lambda|\,\int_0^1\left|F^{-1}_{\dd_{\mu,m}(\mu)}(s)\right|\,\dd s\\
&=|1-\lambda|\,\E_{\mu}\left[\dd_{\mu,m}(X)\right].
\end{align*}
\epv
\subsection{The case of uniform measures on non-empty bounded open subsets of $\R^d$}

The DTM-signature discriminates between two uniform measures over two non-empty bounded open subsets of $\R^d$ with different Lebesgue volume.

\prop
Let $(O,\|\cdot\|_2,\mu_O)$ and $(O',\|\cdot\|_2,\mu_{O'})$ be two mm-spaces, for $O$ and $O'$ two non-empty bounded open subsets of $\R^d$ satisfying $O=\left(\overline{O}\right)^{\circ}$ and $O'=\left({\overline{O'}}\right)^{\circ}$, and $\|.\|_2$ the euclidean norm.
A lower bound for $W_1(\dd_{\mu_O,m}(\mu_O),\dd_{\mu_{O'},m}(\mu_{O'}))$ is given by:
\[\min\left(\mu_O\left(O_{\epsilon(m,O)}\right),\mu_{O'}\left({O'}_{\epsilon(m,O')}\right)\right)\frac{d}{d+1}\left(\frac{m}{\omega_d}\right)^\frac1d\left|\Leb_d(O)^\frac{1}{d}-\Leb_d(O')^\frac{1}{d}\right|.\]
Here,
$O_\epsilon=\left\{x\in O\,|\,\dd(x,\partial O)\geq\epsilon\right\},$
and 
$\epsilon(m,O)=\left(\frac{m\Leb_d(O)}{\omega_d}\right)^\frac1d$ is the radius of any ball of $\mu_O$ mass $m$, included in $O$.
\eprop
\pv
If the set $O_{\epsilon(m,O)}$ is non-empty, then the minimal value of the distance to a measure is given by:
\[\min_{x\in\R^d}(\dd_{\mu_O,m}(x))=\dd_{\min}:=\frac{d}{d+1}\left(\frac{m\Leb_d(O)}{\omega_d}\right)^\frac1d.\]
Moreover, the points at minimal distance are exactly the points of $O_{\epsilon(m,O)}$.
This is Proposition \ref{min dtm} in the Appendix.
So, $F_{\dd_{\mu_O,m}(\mu_O)}(\dd_{\min})=\mu_O\left({O}_{\epsilon(m,O)}\right)$. To conclude, we use the definition of the $L_1$-Wasserstein distance as the $L_1$-norm between the cumulative distribution functions.
\epv

\subsection{The case of two measures on the same open subset of $\R^d$ with one measure uniform}
\label{Unif non Unif}

Let $(O,\|\cdot\|_2,\mu_O)$ and $(O,\|\cdot\|_2,\nu)$ be two mm-spaces with $O$ a non-empty bounded open subset of $\R^d$ and $\nu$ a measure absolutely continuous with respect to $\mu_O$. Thanks to the Radon-Nikodym theorem, there is some $\mu_O$-measurable function $f$ on $O$ such that for all Borel set $A$ in $O$:
\[\nu(A)=\int_Af(\omega)\dd\mu_O(\omega).\]
We can consider the $\lambda$-super-level sets of the function $f$ denoted by $\{f\geq\lambda\}$. As for the previous part, we will denote by $\{f\geq\lambda\}_{\epsilon}$ the set of points belonging to $\{f\geq\lambda\}$ whose distance to $\partial\{f\geq\lambda\}$ is at least $\epsilon$.

Then we get the following lower bound for the $L_1$-Wasserstein distance between the two signatures: 
\prop
\label{Borne inf unif}
Under these hypotheses, a lower bound for $W_1(\dd_{\mu_O,m}(\mu_O),\dd_{\nu,m}(\nu))$ is given by:
\[\frac{1}{1+d}\frac{1}{\Leb_d(O)}\left(\frac{m\Leb_d(O)}{\omega_d}\right)^\frac1d\int_{\lambda=1}^\infty\frac{1}{\lambda^\frac1d}\max_{\lambda'\geq\lambda} \Leb_d\left(\{f\geq\lambda'\}_{\left(\frac{m}{\omega_d}\frac{\Leb_d(O)}{\lambda'}\right)^\frac1d}\right)\dd\lambda.\]
\eprop
\pv
Proof in the Appendix, in Section \ref{B2}.\epv

\subsubsection*{When the density $f$ is H\"older}

We assume that $f$ is H{\"o}lder on $O$, with positive parameters $\chi\in(0,1]$ and $L>0$, that is:
\[\forall x,y\in O,\ |f(x)-f(y)|\leq L\|x-y\|_2^\chi.\]
We also assume that $\Reach(O)>0$. Then for $m$ small enough, the DTM-signature is discriminative.

\prop
\label{Dis Holder}
Under the previous assumptions, if one of the following conditions is satisfied, then the quantity $W_1(\dd_{\mu_O,m}(\mu_O),\dd_{\nu,m}(\nu))$ is positive:
\[m<\frac{\omega_d}{\Leb_d(O)}\min\left\{\Reach(O)^d,\left(\frac{\|f\|_{\infty,O}-1}{2L}\right)^{\frac{d}{\chi}}\right\};\]
\[m\in\left[\frac{\omega_d}{\Leb_d(O)}(\Reach(O))^d,\left(\|f\|_{\infty,O}-2L(\Reach(O))^\chi\right)(\Reach(O))^d\frac{\omega_d}{\Leb_d(O)}\right);\]
\[m\in\left[\frac{\omega_d}{\Leb_d(O)}\left(\frac{d}{\chi}\right)^{\frac{d}{\chi}}(2L)^{-\frac{d}{\chi}},\min\left\{m_0,\frac{\omega_d}{\Leb_d(O)}(\Reach(O))^{d+\chi}\frac{\chi}{d}2L\right\}\right),\]
with $m_0=\|f\|_{\infty,O}^{\frac{d}{\chi}+1}\ \frac{\omega_d}{\Leb_d(O)}\left(\frac{d}{\chi}\right)^{\frac{d}{\chi}}(2L)^{-\frac{d}{\chi}}\left(\frac{\chi}{d+\chi}\right)^{\frac{\chi}{d+\chi}}$.

Moreover, under any of these conditions, we get the lower bound for the quantity $W_1(\dd_{\mu_O,m}(\mu_O),\dd_{\nu,m}(\nu))$:
\[\frac{1}{1+d}\left(\frac{m \Leb_d(O)}{\omega_d}\right)^\frac{1}{d}\int_{\lambda=1}^\infty \frac{1}{\lambda^{1+\frac1d}}\sup_{\lambda'\geq\lambda}\nu\left(\left\{f\geq\lambda'+L\epsilon(\lambda')^\chi\right\}\cap O_{\epsilon(\lambda')}\right)\dd\lambda,\]
with $\epsilon(\lambda')=\lambda'^{-\frac{1}{d}}\left(\frac{m\Leb_d(O)}{\omega_d}\right)^{\frac{1}{d}}$.
\eprop
\pv
Proof in the Appendix, in Section \ref{B2}.
\epv

The previous examples provide several relevant cases where the DTM-signature turns out to be discriminative. It is thus appealing to use it as a tool to compare mm-spaces up to isomorphism.

\section{An algorithm to compare metric-measure spaces from samples}
\label{An algorithm to compare metric-measure spaces from samples}
In this section, $(\x,\delta,\mu)$ and $(\y,\gamma,\nu)$ are two mm-spaces.
We build a test of the null hypothesis
\[H_0\text{ ``The mm-spaces }(\x,\delta,\mu)\text{ and }(\y,\gamma,\nu)\text{ are isomorphic''},\]
against its alternative:
\[H_1\text{ ``The mm-spaces }(\x,\delta,\mu)\text{ and }(\y,\gamma,\nu)\text{ are not isomorphic''}.\]

\subsection{The algorithm}
\label{algo}

The test we propose is based on the fact that the DTM-signatures associated to two isomorphic mm-spaces are equal. If so, it leads to a pseudo-distance  $W_1\left(\dd_{\mu,m}(\mu),\dd_{\nu,m}(\nu)\right)$ equal to zero.

Let consider, in this part, a $N$-sample $P$ from the measure $\mu$, and a $N$-sample $Q$ from the measure $\nu$. A natural idea for a test is to approximate the pseudo-distance by the statistic $W_1\left(\dd_{\1_{P},m}(\1_{P}),\dd_{\1_{Q},m}(\1_{Q})\right)$, where $\1_{P}$ is the uniform probability measure on the set $P$, and to reject the hypothesis $H_0$ if this statistic is larger than some critical value. The choice of the critical value should rely on some parameter $\alpha\in(0,1)$ and lead to a level $\alpha$ for the test. It strongly depends on the measures $\mu$ and $\nu$ that are unknown. Nonetheless, there exist classical ways of approximating a critical value, one is to mimic the distribution of the statistic by replacing the distribution $\mu$ with the distribution $\1_P$ and $\nu$ with $\1_Q$. Unfortunately, this standard method known as \textbf{bootstrap} fails theoretically and experimentally for our framework.

Thus, we propose another kind of bootstrap. For this purpose, we need to take $P'$ a subset of $P$ and $Q'$ a subset of $Q$. The statistic we focus on is $W_1\left(\dd_{\1_{P},m}(\1_{P'}),\dd_{\1_{Q},m}(\1_{Q'})\right)$. It turns out that in this case, the critical value associated to this statistic can be well approximated from the samples $P$ and $Q$, for a suitable size of $P'$ and $Q'$ with respect to $N$. \\

This approach leads to the following algorithm. \\


\begin{lstlisting}[frame=single,caption={Test Procedure},label=list:8-6,abovecaptionskip=-\medskipamount]
Input:%*$P$ and $Q$ $N$-samples from $\mu$ (respectively $\nu$), $N$, $n$, $m$, $\alpha$, $N_{MC}$ even*);
# Compute %*$T$*) the test statistic
%*Take $P'$ a random subset of $P$ of size $n$*);
%*Take $Q'$ a random subset of $Q$ of size $n$*);
%*$T\leftarrow \sqrt{n}W_1(\dd_{\1_{P},m}(\1_{P'}),\dd_{\1_{Q},m}(\1_{Q'}))$*);
# Compute %*$boot$*) a %*$N_{MC}$*)-sample from the bootstrap law
%*$dtmP\leftarrow (\dd_{\1_{P},m}(x))_{x\in P}$*);
%*$dtmQ\leftarrow (\dd_{\1_{Q},m}(x))_{x\in Q}$*);
%*Let $boot$ be empty*);
for j in 1..%*$\lfloor N_{MC}/2\rfloor$*):
  %*Let $dtmP_1$ and $dtmP_2$ be two independent $n$-samples from $\1_{dtmP}$*);
  %*Let $dtmQ_1$ and $dtmQ_2$ be two independent $n$-samples from $\1_{dtmQ}$*);
  %*Add $\sqrt{n}W_1(\1_{dtmP_1},\1_{dtmP_2})$ and $\sqrt{n}W_1(\1_{dtmQ_1},\1_{dtmQ_2})$ to $boot$*);
# Compute %*$qalph$*), the %*$\alpha$*)-quantile of %*$boot$*)
%*Let $qalph$ be the $\lfloor N_{MC}-N_{MC}\times\alpha \rfloor$th smallest element of $boot$*);
Output:%*$(T\geq qalph)$*)
\end{lstlisting}

Recall that the $L_1$-Wasserstein distance $W_1$ is simply the $L_1$-norm of the difference between the cumulative distribution functions. It can be implemented by the $\mathtt{R}$ function $\mathtt{emdw}$ from the package $\mathtt{emdist}$. To compute the distance to an empirical measure at a point $x$, it is sufficient to search for its nearest neighbours; see section \ref{The distance to a measure to dicriminate between measures}. This can be implemented by the $\mathtt{R}$ function $\mathtt{dtm}$ with tuning parameter $r=1$, from the package $\mathtt{TDA}$ \cite{Fasy}.

\subsection{Validity of the method}

In order to prove the validity of our method, we need to introduce a statistical framework.

First of all, from two $N$-samples from the mm-spaces $(\x,\delta,\mu)$ and $(\y,\gamma,\nu)$, we derive four independent empirical measures, $\hat\mu_n$, $\hat\mu_{N-n}$, $\hat\nu_n$ and $\hat\nu_{N-n}$. We also denote $\hat\mu_N$ (respectively $\hat\nu_N$) the empirical measure associated to the whole $N$-sample of law $\mu$ (respectively $\nu$), that is $\hat\mu_N=\frac{n}{N}\hat\mu_n+\frac{N-n}{N}\hat\mu_{N-n}$.

Then, we define the \textbf{test statistic} as:
\[T_{N,n,m}(\mu,\nu)=\sqrt{n}W_1\left(\dd_{\hat\mu_{N},m}(\hat\mu_n),\dd_{\hat\nu_{N},m}(\hat\nu_n)\right).\]
Its law will be denoted by $\LLL_{N,n,m}(\mu,\nu)$.

Remark that for two isomorphic mm-spaces $(\x,\delta,\mu)$ and $(\y,\gamma,\nu)$, the distribution of $T_{N,n,m}(\mu,\nu)$ is $\LLL_{N,n,m}(\mu,\mu)$, $\LLL_{N,n,m}(\nu,\nu)$, but also $\frac12\LLL_{N,n,m}(\mu,\mu)+\frac12\LLL_{N,n,m}(\nu,\nu)$; see Lemma \ref{Invariance lois sous H0} in the Appendix.

For some $\alpha>0$, we denote by $\qu_{\alpha}=\inf\{x\in\R\,|\,F(x)\geq1-\alpha\}$, the \textbf{$\alpha$-quantile} of a distribution with cumulative distribution function $F$ .

The $\alpha$-quantile $\qu_{\alpha,N,n}$ of $\frac12\LLL_{N,n,m}(\mu,\mu)+\frac12\LLL_{N,n,m}(\nu,\nu)$ will be approximated by the $\alpha$-quantile $\hat\qu_{\alpha,N,n}$ of $\frac12\LLL^*_{N,n,m}(\hat\mu_N,\hat\mu_N)+\frac12\LLL^*_{N,n,m}(\hat\nu_N,\hat\nu_N)$. Here $\LLL^*_{N,n,m}(\hat\mu_N,\hat\mu_N)$ stands for the distribution of
$\sqrt{n}W_1\left(\dd_{\hat\mu_N,m}(\mu^*_n),\dd_{\hat\mu_N,m}(\mu'^*_n)\right)$
conditionally to $\hat\mu_N$, where $\mu^*_n$ and $\mu'^*_n$ are two empirical measures from independent $n$-samples of law $\hat\mu_N$.

The \textbf{test} we deal with in this paper is then:
\[\phi_N=\1_{T_{N,n,m}(\mu,\nu)\geq\hat\qu_{\alpha,N,n}}.\]

The null hypothesis $H_0$ is rejected if $\phi_N=1$, that is if the $L_1$-Wasserstein distance between the two empirical signatures $\dd_{\hat\mu_{N},m}(\hat\mu_n)$ and $\dd_{\hat\nu_{N},m}(\hat\nu_n)$ is too high.

\subsubsection{A test of asymptotic level $\alpha$}

In this part, we prove that the test we propose is of asymptotic level $\alpha$, that is such that:
\[\limsup_{N\rightarrow\infty}\p_{(\mu,\nu)\in H_0}(\phi_N=1)\leq\alpha.\]
For this, we prove that the law of the test statistic $\frac12\LLL_{N,n,m}(\mu,\mu)+\frac12\LLL_{N,n,m}(\nu,\nu)$ under the hypothesis $H_0$ and the bootstrap law $\frac12\LLL^*_{N,n,m}(\hat\mu_N,\hat\mu_N)+\frac12\LLL^*_{N,n,m}(\hat\nu_N,\hat\nu_N)$ converge weakly to some fixed distribution when $n$ and $N$ go to $\infty$. In order to adopt a non-asymptotic and more visual point of view, we also derive upper bounds in expectation for the $L_1$-Wasserstein distance between these two distributions.

Remark that it is sufficient to prove weak convergence for $\LLL_{N,n,m}(\mu,\mu)$ and $\LLL^*_{N,n,m}(\hat\mu_N,\hat\mu_N)$. Moreover,
\[W_1\left(\frac12\LLL_{N,n,m}(\mu,\mu)+\frac12\LLL_{N,n,m}(\nu,\nu),\frac12\LLL^*_{N,n,m}(\hat\mu_N,\hat\mu_N)+\frac12\LLL^*_{N,n,m}(\hat\nu_N,\hat\nu_N)\right)\]
is upper bounded by
\[\frac{1}{2}W_1\left(\LLL_{N,n,m}(\mu,\mu),\LLL^*_{N,n,m}(\hat\mu_N,\hat\mu_N)\right)+\frac{1}{2}W_1\left(\LLL_{N,n,m}(\nu,\nu),\LLL^*_{N,n,m}(\hat\nu_N,\hat\nu_N)\right).\]

 This is a straightforward consequence of the definition of the $L_1$-Wasserstein distance with transport plans. Thus, this is also sufficient to derive upper bounds in expectation for the quantity $W_1\left(\LLL_{N,n,m}(\mu,\mu),\LLL^*_{N,n,m}(\hat\mu_N,\hat\mu_N)\right)$.
\lm
\label{Asymptotic law to bridge}
For $\mu$ a measure supported on a compact set, we choose $n$ as a function of $N$ such that: when $N$ goes to infinity, $n$ goes to infinity, $\sqrt{n}\E[\|\dd_{\mu,m}-\dd_{\hat\mu_{N},m}\|_{\infty,\x}]$ goes to zero or more specifically $\frac{\sqrt{n}}{m}\E[W_1(\mu,\hat\mu_{N})]$ goes to zero. Then we have that:
\[\LLL_{N,n,m}(\mu,\mu)\leadsto\LLL\left(\|\G_{\mu,m}-\G'_{\mu,m}\|_1\right),\]
when $N$ goes to infinity.
Moreover, if $n$ is chosen such that $\sqrt{n}W_1(\dd_{\mu,m}(\mu),\dd_{\mu,m}(\hat\mu_N))$ and $\sqrt{n}\|\dd_{\mu,m}-\dd_{\hat\mu_N,m}\|_{\infty,\x}$ go to zero a.e., we have that for almost every sample $X_1, X_2,\ldots X_N\ldots$:
\[\LLL^*_{N,n,m}(\hat\mu_N,\hat\mu_N)\leadsto\LLL\left(\|\G_{\mu,m}-\G'_{\mu,m}\|_1\right),\]
when $N$ goes to infinity; with $\G_{\mu,m}$ and $\G'_{\mu,m}$ two independent Gaussian processes with covariance kernel $\kappa(s,t)=F_{\dd_{\mu,m}(\mu)}(s)\left(1-F_{\dd_{\mu,m}(\mu)}(t)\right)$ for $s\leq t$.
\elm
\pv Proof in the Appendix, in Section \ref{Conv loi pont brownien}.
\epv

\prop
If the two weak convergences in lemma \ref{Asymptotic law to bridge} occur, and if the $\alpha$-quantile $\qu_\alpha$ of the distribution $\LLL(\frac12\|\G_{\mu,m}-\G'_{\mu,m}\|_1+\frac12\|\G_{\nu,m}-\G'_{\nu,m}\|_1)$ is a point of continuity of its cumulative distribution function, then the asymptotic level of the test at $(\mu,\nu)$ is $\alpha$.
\eprop
\pv Proof in the Appendix, in Section \ref{Conv loi pont brownien}.
\epv
Remark that for uniform measures on any sphere in $\R^d$, the continuity assumption for the cumulative distribution function of $\LLL(\|\G_{\mu,m}-\G'_{\mu,m}\|_1)$ is not satisfied. This is a degenerated case. Thus, the test cannot be applied to such mm-spaces.\\

We choose $N=cn^\rho$ for some positive constants $\rho$ and $c$. Then the test is asymptotically valid for two measures supported on a compact subset of the Euclidean space $\R^d$ if we assume that $\rho>\frac{\max\{d,2\}}{2}$.

\prop
\label{prop compact}
Let $\mu$ be some Borel probability measure supported on some compact subset of $\R^d$. Under the assumption
\[\rho>\frac{\max\{d,2\}}{2},\]
the two weak convergences of lemma \ref{Asymptotic law to bridge} occur.

Moreover, a bound for the expectation of $W_1\left(\LLL_{N,n,m}(\mu,\mu),\LLL^*_{N,n,m}(\hat\mu_N,\hat\mu_N)\right)$ is of order:
\[N^{\frac{1}{2\rho}-\frac{1}{\max\{d,2\}}}\left(\log(1+N)\right)^{\1_{d=2}}.\]
And, $W_1\left(\LLL_{N,n,m}(\mu,\mu),\LLL^*_{N,n,m}(\hat\mu_N,\hat\mu_N)\right)\rightarrow0$ a.e. when $n$ goes to $\infty$.
\eprop
\pv This proposition is based on rates of convergence for the Wasserstein distance between a measure $\mu$ with values in $\R^d$ and its empirical version $\hat\mu_N$; see \cite{Fournier} for general dimensions and \cite{Bobkov} for $d=1$. Proof in the Appendix, in Section \ref{section compact supported}.
\epv

A probability measure $\mu$ is \textbf{$(a,b)$-standard} with positive parameters $a$ and $b$, if for all positive radius $r$ and any point $x$ of the support of $\mu$, we have that $\mu(\B(x,r))\geq \min\{1,ar^b\}$. Uniform measures on open subsets of $\R^d$ satisfy such a property:

\ex
Let $O$ be a non-empty bounded open subset of $\R^d$. Then, the measure $\mu_O$ is $(a,d)$-standard with \[a=\frac{\omega_d}{\Leb_d(O)}\left(\frac{\Reach(O)}{\Diam(O)}\right)^d.\]
Here, $\Diam(O)$ stands for the diameter of $O$ and $\omega_d$ for $\Leb_d(\B(0,1))$, the Lebesgue volume of the unit $d$-dimensional ball.
\eex
\pv Proof in the Appendix, in Section \ref{B1}.
\epv

Similar results can be obtained for uniform measures on compact submanifolds of dimension $d$. In \cite{Niyogi} (lemma 5.3), the authors give a bound for $a$ depending on the reach of the submanifold.

The test is asymptotically valid for two $(a,b)$-standard measures supported on compact connected subsets of $\R^d$ if $\rho>1$:

\prop
\label{prop std}
Let $\mu$ be an $(a,b)$-standard measure supported on a connected compact subset of $\R^d$. The two weak convergences of lemma \ref{Asymptotic law to bridge} occur if the assumption $\rho>1$ is satisfied.
Moreover, a bound for the expectation of $W_1\left(\LLL_{N,n,m}(\mu,\mu),\LLL^*_{N,n,m}(\hat\mu_N,\hat\mu_N)\right)$ is of order $N^{\frac{1}{2\rho}-\frac{1}{2}}$ up to a logarithm term.
\eprop
\pv This proposition is based on rates of convergence for the infinity norm between the distance to a measure and its empirical version; see \cite{Chazal}. Proof in the Appendix, in Section \ref{C3}.
\epv

Remark that we can achieve a rate close to the parametric rate for Ahlfors regular measures, whereas for general measures, the rate gets worse when the dimension increases. Anyway, we need $\rho$ to be as big as possible for the bootstrapped law to be a good enough approximation of the law of the statistic, that is to have a type I error close enough to $\alpha$; keeping in mind that $n$ should go to $\infty$ with $N$.

\subsubsection{The power of the test}
The \textbf{power} of the test $\phi_N=\1_{\sqrt{n}W_1\left(\dd_{\hat\mu_{N},m}(\hat\mu_n),\dd_{\hat\nu_{N},m}(\hat\nu_n)\right)\geq\hat\qu_{\alpha,N,n}}$ is defined for two mm-spaces $(\x,\delta,\mu)$ and $(\y,\gamma,\nu)$ by:
\[1-\p_{(\mu,\nu)}\left(\phi_N=0\right).\]
If the spaces are not isomorphic, we want the test to reject the null with high probability. It means that we want the power to be as big as possible.
Here, we give a lower bound for the power, or more precisely an upper bound for $\p_{(\mu,\nu)}\left(\phi_N=0\right)$, the \textbf{type II error}.
\prop
\label{Power du test}
Let $\mu$ and $\nu$ be two Borel measures supported on $\x$ and $\y$, two compact subsets of $\R^d$. We assume that the mm-spaces $(\x,\delta,\mu)$ and $(\y,\gamma,\nu)$ are non-isomorphic and that the DTM-signature is discriminative for some $m$ in $(0,1]$, that is such that $W_1\left(\dd_{\mu,m}(\mu),\dd_{\nu,m}(\nu)\right)>0$. We choose $N=n^\rho$ with $\rho>1$.
Then for all positive $\epsilon$, there exists $n_0$ depending on $\mu$ and $\nu$ such that for all $n\geq n_0$, the type II error \[\p_{(\mu,\nu)}\left(\sqrt{n}W_1\left(\dd_{\hat\mu_{N},m}(\hat\mu_n),\dd_{\hat\nu_{N},m}(\hat\nu_n)\right)<\hat\qu_{\alpha,N,n}\right)\] is upper bounded by \[4\exp\left(-\frac{W_1^2\left(\dd_{\mu,m}(\mu),\dd_{\nu,m}(\nu)\right)}{(2+\epsilon)\max\left\{\Diam_{\mu,m}^2,\Diam_{\nu,m}^2\right\}}n\right),\]
with $\Diam_{\mu,m}$, the diameter of the support of the measure $\dd_{\mu,m}(\mu)$.
\eprop
\pv
Proof in the Appendix, in Section \ref{C4}.
\epv
In order to have a high power, that is to reject $H_0$ more often when the mm-spaces are not isomorphic, we need $n$ to be big enough, that is $\rho$ small enough. Recall that $n$ has to be small enough for the law of the statistic and its bootstrap version to be close. It means that some compromise should be done. Moreover, the choice of $m$ for the test should depend on the geometry of the mm-spaces. The tuning of these parameters from data is still an open question.

\section{Numerical illustrations}
\label{illustration}
Let $\mu_{v}$ be the distribution of the random vector $(R\sin(v R)+0.03 N,R\cos(v R)+0.03 N')$ with $R$, $N$ and $N'$ independent random variables; $N$ and $N'$ from the standard normal distribution and $R$ uniform on $(0,1)$.
With the notation given in the Introduction, we consider the sets $\Theta_0=\{(\mu_{10},\mu_{10})\}$ and $\Theta_1=\{(\mu_{10},\mu_{p})\,|\,p\neq10\}$.
We sample $N=2000$ points from two measure, choose $\alpha=0.05$, $m=0.05$, $n=20$, and $N_{MC}=1000$. We give an example under which our test (\textbf{DTM}) is working and more powerful than (\textbf{KS}), which consists in applying a Kolmogorov-Smirnov test to $\frac{N}{2}$-samples from $\LLL(\delta(X,X'))$ and $\LLL(\gamma(Y,Y'))$ with  $X$ and $X'$ (resp. $Y$ and $Y'$) independent from $\mu$ (resp. $\nu$). The experiments are repeated 1000 times to approximate the type I error for our test and the power for both tests.

\begin{figure}[h]
 \begin{minipage}[h]{.49\linewidth}
  \centering\includegraphics[clip=true,trim=0cm 1.7cm 0cm 2cm,scale=0.42]{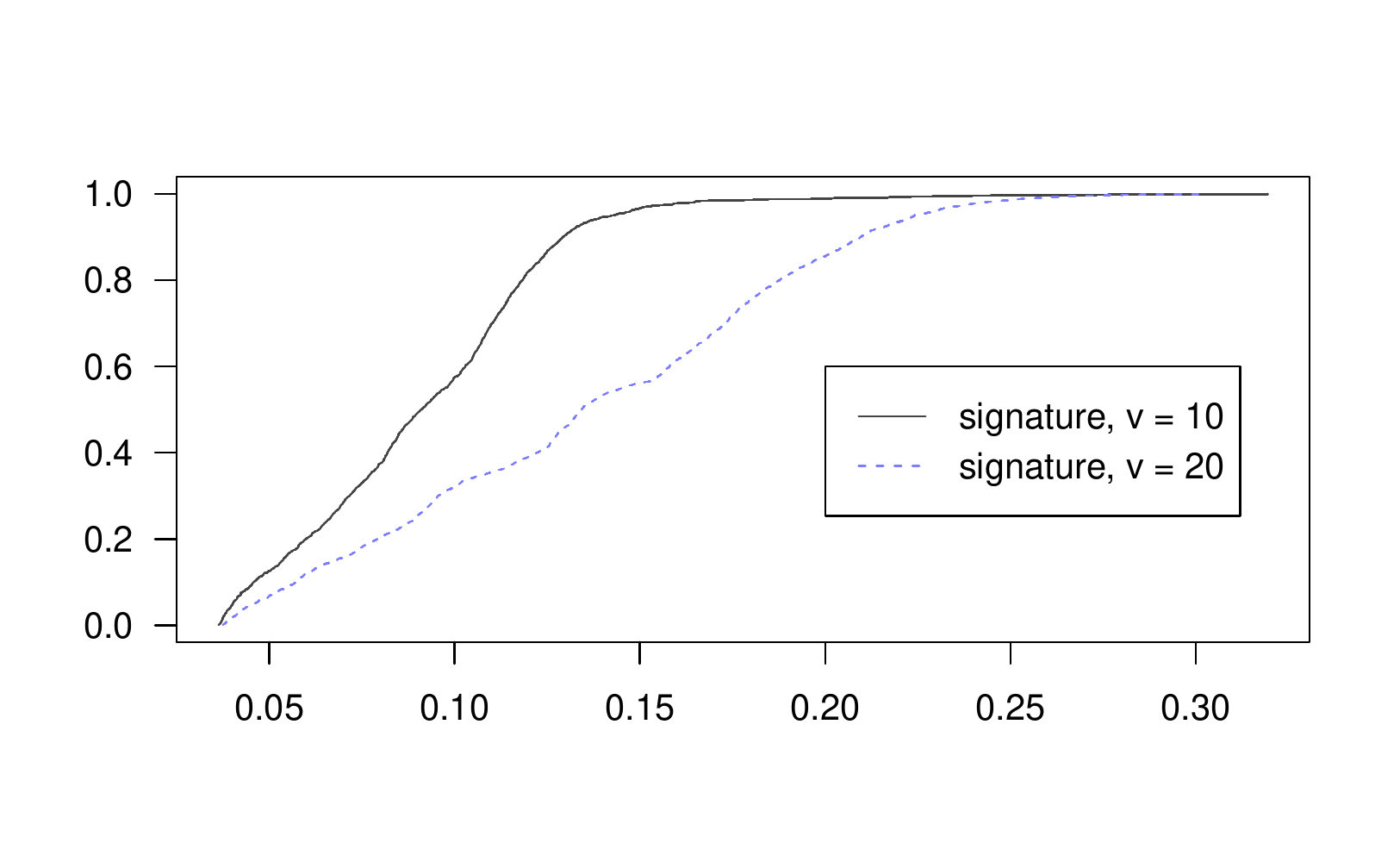}
  \caption{\label{signaturem005} DTM-signature estimates, $m = 0.05$}
 \end{minipage}
  \begin{minipage}[h]{.49\linewidth}
  \centering\includegraphics[clip=true,trim=0cm 1.7cm 0cm 2cm,scale=0.42]{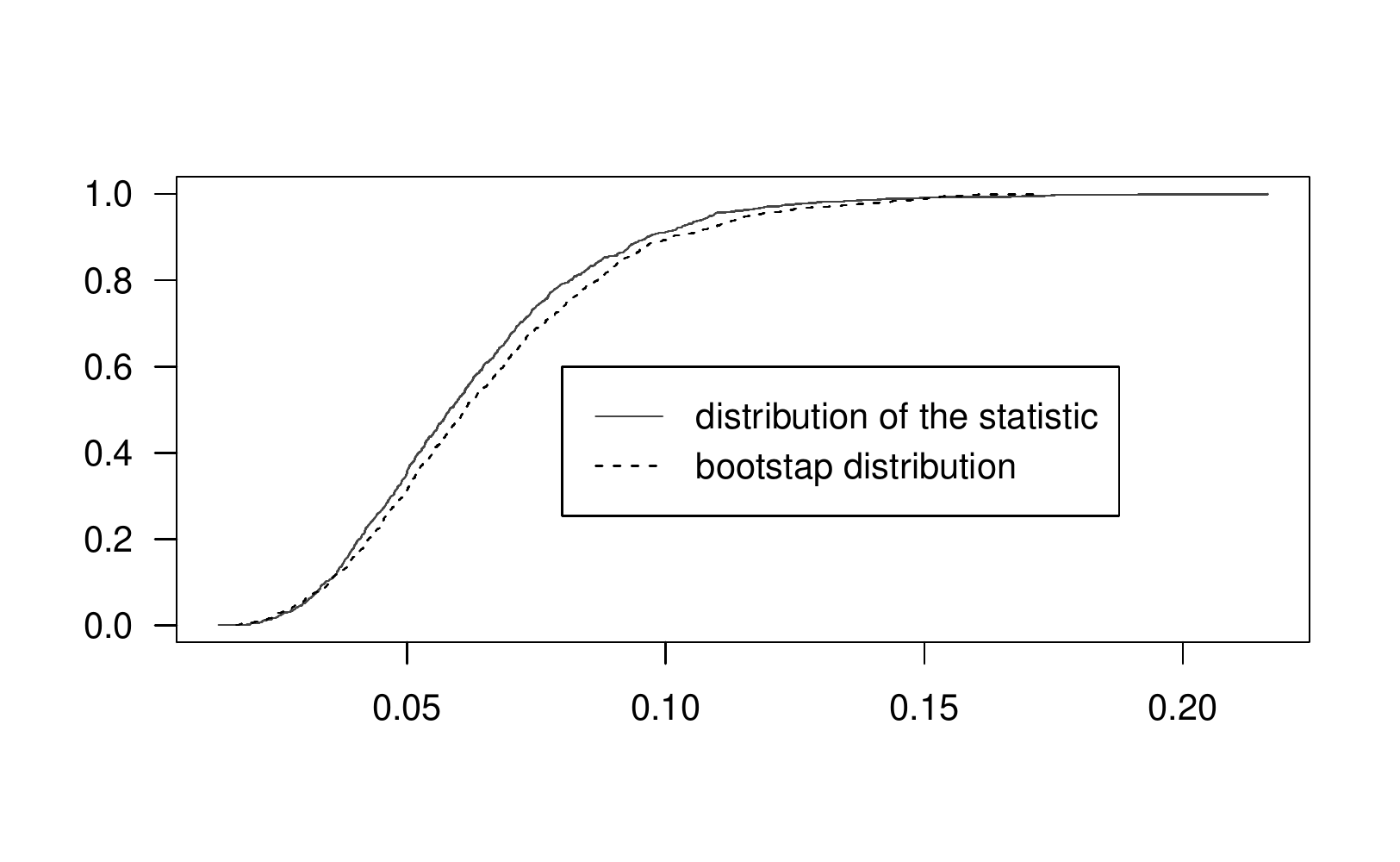}
  \caption{\label{Bootstrapm005} Bootstrap validity, $v = 10$, $m = 0.05$}
 \end{minipage}
\end{figure}

\begin{figure}[h]
 \begin{minipage}[h]{.69\linewidth}
\begin{tabular}{l c c c c c}
\hline
v & 15 & 20 & 30 & 40 & 100\\
\hline
\rowcolor{Gray}
\cellcolor{white}type I error \textbf{DTM} & 0.050 & 0.049 & 0.051 & 0.044 & 0.051 \\
power \textbf{DTM} & 0.525 & 0.884 & 0.987 & 0.977 & 0.985\\
\rowcolor{Gray}
\cellcolor{white}power \textbf{KS} & 0.768 & 0.402 & 0.465 & 0.414 & 0.422\\
\hline
\end{tabular}
\caption{\label{table}Type I error and power approximations}
 \end{minipage}
  \begin{minipage}[h]{.3\linewidth}
  \centering\includegraphics[clip=true,trim=0cm 2cm 0cm 2cm,scale=0.13]{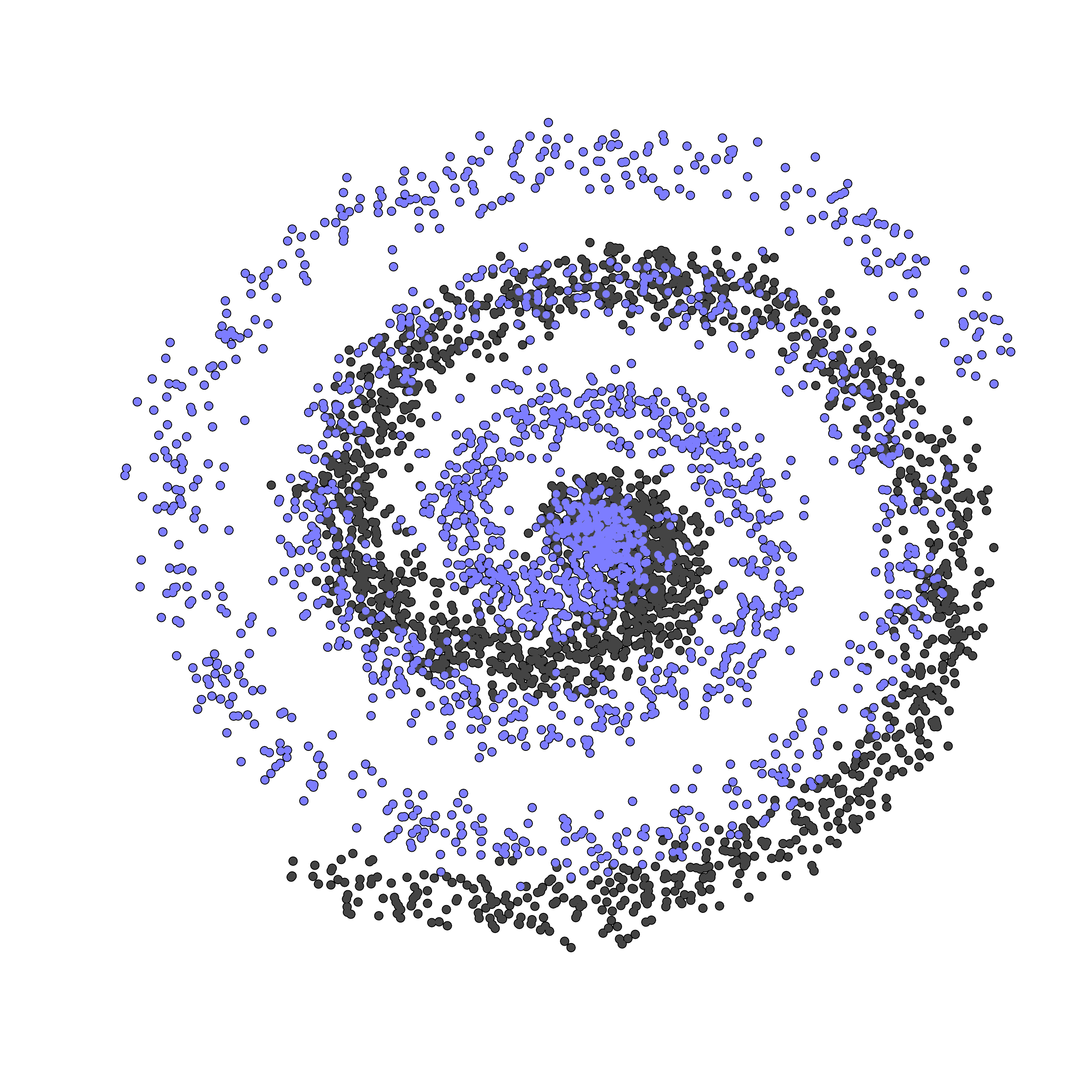}
 \end{minipage}
\end{figure}

\section{Concluding remarks and perspectives}
This paper opens a new horizon of statistical tests based on shape signatures. It could be of interest to adapt these kind of methods to other signatures, if possible. In future it could even be interesting to build statistical tests based on many different signatures, leading to an even better discrimination. Regarding the test proposed in this paper itself, the geometric and statistical problem of the choice of the best parameters to use in practice is still an open, tough and engaging question.

\subparagraph*{Acknowledgements}
The author is extremely grateful to Fr\'ed\'eric Chazal, Pascal Massart and Bertrand Michel for introducing her to the distance to a measure, for their valuable comments and advises, and for proofreading.

\printbibliography

\pagebreak
\part*{Appendix}
\begin{appendix}
\section{Uniform measures on open subsets of $\R^d$}

In this part, we focus on some mm-spaces $(O,\|\cdot\|_2,\mu_O)$ where $O$ stands for a non-empty bounded open subset of $\R^d$ satisfying $\left(\overline{O}\right)^\circ=O$. The measure $\mu_O$, the medial axis $\MM(O)$ and the reach $\Reach(O)$ have been defined in Section \ref{Uniform measures on non-empty open subsets}. The object $\epsilon(m,O)$ is defined for some mass parameter $m$ in $[0,1]$ by 
\[\epsilon(m,O)=\left(\frac{m\Leb_d(O)}{\omega_d}\right)^\frac1d.\] This is the radius of a ball included in $O$, with $\mu_O$ measure equal to $m$.
For some positive $\epsilon$, $O_\epsilon$ stands for the set of points in $O$ which distance to $\partial O$ is not smaller than $\epsilon$:
\[O_\epsilon=\left\{x\in O, \inf_{y\in\partial O}\|x-y\|_2\geq\epsilon\right\}.\]

\subsection{The distance to uniform measures}
\label{B1}
Here, we derive some properties of the spaces $(O,\|\cdot\|_2,\mu_O)$. We give a lower bound for the minimum of the distance to the measure $\mu_O$ and give a description of the points attaining this bound. Then, we use such considerations to prove identifiability of the measure $\mu_O$ from its distance-to-a-measure function. That is, to prove Proposition \ref{inject measure} of the paper.

First, we state some technical lemma proposed by Lieutier in \cite{Lieutier}.

\lm
\label{medial axis and skeleton}
If we define the \textbf{skeleton} $\Sk(O)$ of the open set $O$ as the set of centres of maximal balls (for the inclusion) included in $O$, then we get:
\[\MM(O)\subset\Sk(O)\subset\overline{\MM(O)}.\]
\elm

Now we can formulate some technical lemma:

\lm
\label{Inclusion_boule_maximale}
For any $x$ in $O$, there exist a maximal ball for the inclusion, included in $O$ and containing $x$.
\elm
\pv
Let us consider the class $\SSS=\{\B(y,r)\,|\,r>0\text{ and }x\in\B(y,r)\subset O\}$ of all non-empty open balls included in $O$ and containing $x$. We are going to show that this class contains a maximal element by using the Zorn's lemma. For this, we need to show that the partially-ordered set $\SSS$ is inductive, which means that any non-empty totally-ordered subclass $\T$ of $\SSS$ is upper bounded by some element of $\SSS$.
Let $\T$ be a non-empty totally-ordered subclass of $\SSS$. Set $R=\sup\{r>0\,|\,\exists\,y\in O, \B(y,r)\in\T\}$ the supremum of the radii of all balls in $\T$. Since $\T$ is non-empty and $O$ is bounded, $R$ if positive and finite.
Let $(y_k)_{k\in\N}$ be a sequence of centres of balls in $\T$ converging to a point $y$ in $\R^d$ such that the sequence of associated radii $(r_k)_{k\in\N}$ is non decreasing with $R$ as a limit.
Since $\T$ is totally-ordered and the radii non decreasing, the union $\bigcup_{k\in\N}\B(y_k,r_k)$ is non decreasing, equal to $\B(y,R)$. Thus, $\B(y,R)$ belongs to $\SSS$ and upper bounds $\T$.
So the class $\SSS$ is inductive and thanks to the Zorn's lemma, it contains a maximal element.
\epv

\paragraph*{Proof of Example 19:}
For any point $x$ in $O$ and $r>0$, thanks to Lemma \ref{Inclusion_boule_maximale} there exist a maximal ball $\B(x',r')$ included in $O\cap\B(x,r)$ which contains $x$. Assume for the sake of contradiction that $r'<\min\left\{\frac r2,\Reach(O)\right\}$.

Since $r'<\frac r2$, the ball $\overline{\B}(x',r')$ is included in $\B(x,r)$ thus $\B(x',r')$ is maximal in $O$. So $x'$ belongs to $\Sk(O)$, and thanks to Lemma \ref{medial axis and skeleton}, to $\overline{\MM(O)}$. But $r'<\Reach(O)$; this is absurd.

It follows that:
\[\mu_O(\B(x,r))\geq\mu_O\left(\B\left(x',\min\left\{\Reach(O),\frac r2\right\}\right)\right).\]
So, for $r\leq2\Reach(O)$, since $2\Reach(O)\leq\Diam(O)$ by considering a point on $\Sk(O)$, we get:
\[\mu_O(\B(x,r))\geq r^d \left(\frac{\Reach(O)}{\Diam(O)}\right)^d\frac{\omega_d}{\Leb_d(O)},\]
which is also true for $r$ in $[2\Reach(O),\Diam(O)]$, whereas for $r\geq\Diam(O)$ we have $\mu_O(\B(x,r))=1$.
The choice of $a$ in the lemma is thus relevant.
$\blacksquare$\\

We now focus on the set of points in $\R^d$ minimizing the distance to the measure $\mu_O$. For this, we need some lemma.

\lm
\label{ball in O}
If $x$ in $\R^d$ satisfies $\mu_O(\B(x,\epsilon))=\frac{\omega_d\epsilon^d}{\Leb_d(O)}$, then $\B(x,\epsilon)\subset O$.
\elm

\pv
If $x$ in $\R^d$ satisfies $\mu_O(\B(x,\epsilon))=\frac{\omega_d\epsilon^d}{\Leb_d(O)}$, then, $\Leb_d(O^c\cap\B(x,\epsilon))=0$.
Assume for the sake of contradiction that the set $O^c\cap\B(x,\epsilon)$ is not empty. Since $\left(\overline{O}\right)^\circ=O$, then the open subset $\left(O^c\right)^\circ\cap\B(x,\epsilon)$ of $O^c\cap\B(x,\epsilon)$ is not empty, thus of positive Lebesgue measure, which is absurd.
So $\B(x,\epsilon)\subset O$.
\epv

\prop
\label{min dtm}
The constant $\dd_{\min}=\frac{d}{d+1}\left(\frac{m\Leb_d(O)}{\omega_d}\right)^\frac1d$ is a lower bound for the distance to the measure $\mu_O$ over $\R^d$.
Moreover, the set of points attaining this bound is exactly $O_{\epsilon(m,O)}$.
\eprop

\pv
Remark that for all positive $l$ smaller than $m$, we have: \[\delta_{\mu,l}(x)\geq\left(\frac{l\Leb_d(O)}{\omega_d}\right)^\frac1d.\]
Moreover, these inequalities are equalities for all points $x$ in $O_{\epsilon(m,O)}$.
By integrating, we get the lower bound $\dd_{\min}$ for $x\mapsto\dd_{\mu,m}(x)$, and it is attained on $O_{\epsilon(m,O)}$.

Now take some point $x$ in $\R^d$ satisfying $\dd_{\mu,m}(x)=\dd_{\min}$. For almost all $l$ smaller than $m$, we have: $\delta_{\mu,l}(x)=\left(\frac{l \Leb_d(O)}{\omega_d}\right)^\frac1d$. In particular we get for these values of $l$ that:
\[\mu\left(\overline{\B}\left(x,\left(\frac{m\Leb_d(O)}{\omega_d}\right)^\frac1d\right)\right)>l.\]
So, $\mu\left(\B\left(x,\left(\frac{m\Leb_d(O)}{\omega_d}\right)^\frac1d\right)\right)=m$, and thanks to Lemma \ref{ball in O}, we get that $x\in O_{\epsilon(m,O)}$.
\epv

\prop
\label{id of measures}
If $\Reach(O)\geq\epsilon(m,O)$, then:
\[\{x\in\R^d\,|\,\dd_{\mu,m}(x)=\dd_{\min}\}^{\epsilon(m,O)}=O,\]
where for any set $A$, the notation $A^\epsilon$ stands for $\bigcup_{x\in A}\overline{\B}(x,\epsilon)$, the \textbf{$\epsilon$-offset} of $A$.
\eprop

\pv
Remind thanks to Proposition \ref{min dtm} that $\{x\in\R^d\,|\,\dd_{\mu,m}(x)=\dd_{\min}\}=O_{\epsilon(m,O)}$.
Moreover, $O_{\epsilon(m,O)}^{\epsilon(m,O)}\subset O$.
Assume for the sake of contradiction that the set $O\backslash O_{\epsilon(m,O)}^{\epsilon(m,O)}$ is non-empty. Take a point $x$ in this set and consider $\B(x',r')$ a maximal ball containing $x$ and included in $O$ given by Lemma \ref{Inclusion_boule_maximale}. Since $x\notin O_{\epsilon(m,O)}^{\epsilon(m,O)}$, we get that $r'<\epsilon(m,O)$. Moreover, $x'$ belongs to $\Sk(O)$ and so, thanks to Lemma \ref{medial axis and skeleton}, to $\overline{\MM}(O)$. Then, by continuity of the function distance to the compact set $\partial O$, $r'=\dd_{\partial O}(x')\geq\Reach(O)\geq\epsilon(m,O)$, which is a contradiction. So, $O_{\epsilon(m,O)}^{\epsilon(m,O)}=O$.
\epv

\subsection{The DTM-signature to discriminate between uniform and non uniform measures.}
\label{B2}
\paragraph*{Proof of Proposition \ref{Borne inf unif}:}
As for Proposition \ref{min dtm}, we get that for any point $x$ in $O$:
\[\dd_{\mu_O,m}(x)\geq\dd_{\min}:= \left(\frac{m\Leb_d(O)}{\omega_d}\right)^\frac{1}{d}\frac{d}{1+d}.\]
We will lower bound the $L_1$-Wasserstein distance between $\dd_{\mu_O,m}(\mu_O)$ and $\dd_{\nu,m}(\nu)$ by the integral of $F_{\dd_{\nu,m}(\nu)}$ over the interval $[0,\dd_{\min}]$, since $F_{\dd_{\mu_O,m}(\mu_O)}$ equals zero on this interval.
We thus need to lower bound $F_{\dd_{\nu,m}(\nu)}(t)$ for all $t\leq\dd_{\min}$.

As for Proposition \ref{min dtm}, for $\lambda\geq1$, any point $x$ of $\{f\geq\lambda\}_{\lambda^{-\frac1d}\left(\frac{m\Leb_d(O)}{\omega_d}\right)^{\frac1d}}$ satisfies $\dd_{\nu,m}(x)\leq\frac{\dd_{\min}}{\lambda^\frac1d}$. 
Thus,
\[F_{\dd_{\nu,m}(\nu)}\left(\frac{\dd_{\min}}{\lambda^\frac1d}\right)\geq\nu\left(\{f\geq\lambda\}_{\lambda^{-\frac1d}\left(\frac{m \Leb_d(O)}{\omega_d}\right)^{\frac1d}}\right).\]

And we get by denoting $\lambda(t)$ the real number $\lambda$ satisfying $t=\frac{\dd_{\min}}{\lambda^{\frac{1}{d}}}$, that:
\[W_1(\dd_{\mu_O,m}(\mu_O),\dd_{\nu,m}(\nu))\geq\int_{t=0}^{\dd_{\min}}\nu\left(\{f\geq\lambda(t)\}_{\lambda(t)^{-\frac1d}\left(\frac{m\Leb_d(O)}{\omega_d}\right)^{\frac1d}}\right)\dd t.\]
Since a cumulative distribution function in non decreasing, we get:
\begin{align*}
&W_1(\dd_{\mu_O,m}(\mu_O),\dd_{\nu,m}(\nu))\geq\\
&\int_{t=0}^{\dd_{\min}}\sup_{t'\leq t}\nu\left(\{f\geq\lambda(t')\}_{\lambda(t')^{-\frac1d}\left(\frac{m\Leb_d(O)}{\omega_d}\right)^{\frac1d}}\right)\dd t\\
&=\int_{\lambda=1}^\infty \dd_{\min}\frac{1}{d}\frac{1}{\lambda^{\frac1d}}\frac1\lambda\sup_{\lambda'\geq\lambda}\nu\left(\{f\geq\lambda'\}_{\lambda'^{-\frac1d}\left(\frac{m\Leb_d(O)}{\omega_d}\right)^{\frac1d}}\right)\dd\lambda\\
&\geq \frac{1}{d+1}\left(\frac{m\Leb_d(O)}{\omega_d}\right)^\frac1d\int_{\lambda=1}^\infty\frac{1}{\lambda^{\frac1d}}\sup_{\lambda'\geq\lambda}\mu_O\left(\{f\geq\lambda'\}_{\left(\frac{m\Leb_d(O)}{\lambda'\omega_d}\right)^{\frac1d}}\right)\dd\lambda.
\end{align*}
$\blacksquare$

Now we assume that the density $f$ is H\"older over $O$ with parameters $\chi$ in $[0,1]$ and $L$ in $\R_+^*$.

\paragraph*{Proof of Proposition 15:}
First remark that for all positive $\lambda$, with $\epsilon(\lambda)=\lambda^{-\frac{1}{d}}\left(\frac{m\Leb_d(O)}{\omega_d}\right)^{\frac{1}{d}}$ we have:
\[\left\{f\geq\lambda+L\epsilon(\lambda)^\chi\right\}\cap O_{\epsilon(\lambda)}\subset\{f\geq\lambda\}_{\epsilon(\lambda)}.\]

According to Proposition \ref{Borne inf unif}, the aim is thus to show that for some $\lambda$ bigger than 1, the set $\left\{f\geq\lambda+L\epsilon(\lambda)^\chi\right\}\cap O_{\epsilon(\lambda)}$ is non-empty. We thus focus on the supremum of $f$ over $O_{\epsilon(\lambda)}$, which we denote by $\|f\|_{\infty,\epsilon(\lambda)}$.

Remind that if $\Reach(O)\geq\epsilon(\lambda)$, then thanks to Proposition \ref{id of measures}, the set $O_{\epsilon(\lambda)}^{\epsilon(\lambda)}$ equals $O$. Since $f$ is H\"older, we can thus build some sequence $(y_n)_{n\in\N^*}$ in $O_{\epsilon(\lambda)}$, such that $f(y_n)\geq\|f\|_{\infty,O}-\frac{1}{n}-L\epsilon(\lambda)^\chi$. Finally we get:
\[\|f\|_{\infty,\epsilon(\lambda)}\geq\|f\|_{\infty,O}-L\epsilon(\lambda)^\chi.\]

So the quantity $W_1(\dd_{\mu_O,m}(\mu_O),\dd_{\nu,m}(\nu))$ is positive whenever:
\[\|f\|_{\infty,O}>\inf\left\{\lambda+2L\epsilon(\lambda)^\chi\,|\,\lambda\geq1,\epsilon(\lambda)\leq\Reach(O)\right\}.\]

With $\lambda_0=1$, we have $\lambda_0+2L\epsilon(\lambda_0)^\chi = 1+2L\left(\frac{m\Leb_d(O)}{\omega_d}\right)^{\frac{\chi}{d}}$.

With $\lambda_1$ satisfying $\epsilon(\lambda_1)=\Reach(O)$, we have:
\[\lambda_1+2L\epsilon(\lambda_1)^\chi=\frac{1}{(\Reach(O))^d}\frac{m\Leb_d(O)}{\omega_d}+2L(\Reach(O)^\chi).\]

We also have that
\[\inf\left\{\lambda+2L\epsilon(\lambda)^\chi\,|\,\lambda>0\right\}=(2L)^{\frac{d}{d+\chi}}\left(\frac{\Leb_d(O)}{\omega_d}\right)^\frac{\chi}{d+\chi}m^{\frac{\chi}{d+\chi}}\left[\left(\frac{\chi}{d}\right)^{\frac{d}{\chi+d}}+\left(\frac{\chi}{d}\right)^{-\frac{\chi}{d+\chi}}\right].\]
The infimum is attained at $\lambda_2=\left(\frac{\chi}{d}\right)^{\frac{d}{\chi+d}}(2L)^{\frac{d}{\chi+d}}\left(\frac{m\Leb_d(O)}{\omega_d}\right)^{\frac{\chi}{\chi+d}}$.

It proves the first part of the proposition.

The second part is a straightforward consequence of the proof of Proposition \ref{Borne inf unif}.
$\blacksquare$

\section{Stability of the DTM-signature}
\label{section stab}

\paragraph*{Proof of Proposition 9:}
The proof is relatively similar to the ones given by M{\'e}moli in \cite{Memoli} for other signatures.

For any map plan $\pi$ between $\mu$ and $\nu$ Borel measures on $(\x,\delta)$ and $(\y,\gamma)$, we get:

\begin{align*}
&W_1(\dd_{\mu,m}(\mu),\dd_{\nu,m}(\nu))\leq\\
&\int_{\x\times\y}   \left|\dd_{\mu,m}(x)-\dd_{\nu,m}(y)\right|   \dd\pi(x,y)=\\
&\int_{\x\times\y}   \left|\frac1m\int_0^m\delta_{\mu,l}(x)\dd l-\frac1m\int_0^m\delta_{\nu,l}(y)\dd l\right|   \dd\pi(x,y)\leq\\
&\int_{\x\times\y}   \frac1m\int_0^m\left|\delta_{\mu,l}(x) - \delta_{\nu,l}(y)\right|\dd l\,\dd\pi(x,y)=\\
&\frac 1m\int_{\x\times\y}   \int_0^m\left|\inf\{r>0\,|\,\mu(\overline{\B}(x,r))>l\} - \inf\{r>0\,|\,\nu(\overline{\B}(y,r))>l\}\right|\dd l\,\dd\pi(x,y)  =\\
&\frac 1m\int_{\x\times\y}   \int_0^m\left|\int_0^{+\infty}  \left(\1_{\mu(\overline{\B}(x,r))\leq l}-\1_{\nu(\overline{\B}(y,r))\leq l}\right) \dd r\right|\dd l\,\dd\pi(x,y)\leq \\
&\frac 1m\int_{\x\times\y}   \int_0^{+\infty}\int_0^m \left|\1_{\mu(\overline{\B}(x,r))\leq l}-\1_{\nu(\overline{\B}(y,r))\leq l} \right| \dd l\,\dd r\,\dd\pi(x,y)  \leq \\
&\frac 1m\int_{\x\times\y}   \int_0^{+\infty} \left|\mu(\overline{\B}(x,r))\wedge m- \nu(\overline{\B}(y,r))\wedge m \right| \dd r\,\dd\pi(x,y)   \leq \\
&\frac 1m\int_{\x\times\y}   \int_0^{+\infty} \left|\int_{\x\times\y}\left(\1_{\delta(x,x')\leq r}-\1_{\gamma(y,y')\leq r}\right) \dd\pi(x',y') \right|\wedge m\,\dd r\,\dd\pi(x,y)   \leq  \\
&\frac 1m\int_{\x\times\y}   \int_{\x\times\y}\int_0^{+\infty}  \left|\1_{\delta(x,x')\leq r}-\1_{\gamma(y,y')\leq r}\right|  \dd r\,\dd\pi(x',y')\,\dd\pi(x,y) = \\
&\frac 1m\int_{\x\times\y}   \int_{\x\times\y}  \left|\delta(x,x')-\gamma(y,y')\right|    \dd\pi(x',y')\,\dd\pi(x,y),
\end{align*}
which concludes.

\section{The test}

\subsection{A lemma}

\lm[\sc Equality of empirical signatures under the isomorphic assumption]
\label{Invariance lois sous H0}
If $(\x,\delta,\mu)$ and $(\y,\gamma,\nu)$ are two isomorphic mm-spaces, then the distributions of the random variables 
\[\sqrt{n}W_1(\dd_{\hat\mu_{N},m}(\hat\mu_n),\dd_{\hat\mu'_{N},m}(\hat\mu'_n))\] and \[\sqrt{n}W_1(\dd_{\hat\mu_{N},m}(\hat\mu_n),\dd_{\hat\nu_{N},m}(\hat\nu_n))\] are equal. Here the empirical measures are all independent and the measures $\hat\mu'_{N}$ and $\hat\mu'_n$ are from samples from $\mu$.
\elm

\pv
Remark that for $(X'_1,X'_2,\ldots,X'_N)$ a $N$-sample of law $\mu$ and $\phi$ an isomorphism between $(\x,\delta,\mu)$ and $(\y,\gamma,\nu)$, the tuple $(\phi(X'_1),\phi(X'_2),\ldots,\phi(X'_N))$ is a $N$-sample of law $\nu$. Moreover, $\delta(X'_i,X'_j)=\gamma(\phi(X'_i),\phi(X'_j))$ for all $i$ and $j$ in $[\![1,N]\!]$. It follows that the distances and the nearest neighbours are preserved.

Thus, the distributions of $(\dd_{\hat\mu_{N},m}(X'_i))_{i\in[\![1,n]\!]}$ and $(\dd_{\hat\nu_{N},m}(Y_i))_{i\in[\![1,n]\!]}$ are equal.

The lemma follows from the equality:
\begin{align*}
&W_1(\dd_{\hat\mu_{N},m}(\hat\mu_n),\dd_{\hat\nu_{N},m}(\hat\nu_n))\\
&=\int_{0}^{+\infty}\frac{1}{n}\left|\sum_{i=1}^{n}\1_{\dd_{\hat\mu_{N},m}(X_i)\leq s}-\sum_{i=1}^{n}\1_{\dd_{\hat\nu_{N},m}(Y_i)\leq s}\right|\dd s,
\end{align*}
with $(X_1,X_2,\ldots,X_N)$ a $N$-sample from $\mu$.
\epv

\subsection{$L_1$-Wasserstein distance between the laws of interest}
\label{C1}

\label{section uE}
\lm
\label{validity}
The quantity $W_1\left(\LLL_{N,n,m}(\mu,\mu),\LLL^*_{N,n,m}(\hat\mu_N,\hat\mu_N)\right)$ is upper bounded by:
\[2\sqrt{n}\left(\E[\|\dd_{\hat\mu_{N},m}-\dd_{\mu,m}\|_{\infty,\x}]+W_1(\dd_{\mu,m}(\mu),\dd_{\mu,m}(\hat\mu_N))+\|\dd_{\mu,m}-\dd_{\hat\mu_N,m}\|_{\infty,\x}\right).\]
\elm

\pv
Let $(X_1,X_2, \ldots X_N)$ be a $N$-sample of law $\mu$, and $\hat\mu_N$ the associated empirical measure. We can upper bound the $L_1$-Wasserstein distance between the bootstrap law $\LLL^*(\sqrt{n}W_1(\dd_{\hat\mu_N,m}(\mu^*_n),\dd_{\hat\mu_N,m}(\mu'^*_n))|\hat\mu_N)$ and the law of interest $\LLL(\sqrt{n}W_1(\dd_{\hat\mu_{N},m}(\hat\mu_n),\dd_{\hat\mu'_{N},m}(\hat\mu'_n)))$, by:
\begin{align}
&W_1\left(\LLL\left(\sqrt{n}W_1\left(\dd_{\hat\mu_N,m}\left(\mu^*_n\right),\dd_{\hat\mu_N,m}\left(\mu'^*_n\right)\right)|\hat\mu_N\right),\LLL\left(\sqrt{n}W_1\left(\dd_{\mu,m}\left(\mu^*_n\right),\dd_{\mu,m}\left(\mu'^*_n\right)\right)|\hat\mu_N\right)\right)\label{lign4}\\
&+W_1\left(\LLL\left(\sqrt{n}W_1\left(\dd_{\mu,m}\left(\mu^*_n\right),\dd_{\mu,m}\left(\mu'^*_n\right)\right)|\hat\mu_N\right),\LLL\left(\sqrt{n}W_1\left(\dd_{\mu,m}\left(\hat\mu_n\right),\dd_{\mu,m}\left(\hat\mu'_n\right)\right)\right)\right)\label{lign3}\\
&+W_1\left(\LLL\left(\sqrt{n}W_1\left(\dd_{\mu,m}\left(\hat\mu_n\right),\dd_{\mu,m}\left(\hat\mu'_n\right)\right)\right),\LLL\left(\sqrt{n}W_1\left(\dd_{\hat\mu_{N},m}\left(\hat\mu_n\right),\dd_{\hat\mu'_{N},m}\left(\hat\mu'_n\right)\right)\right)\right)\label{lign2}.
\end{align}

We bound the term \ref{lign4} by:
\[2\sqrt{n}\|\dd_{\mu,m}-\dd_{\hat\mu_N,m}\|_{\infty,\x}.\]
 the term \ref{lign3} by 
\[2\sqrt{n}W_1\left(\dd_{\mu,m}(\mu),\dd_{\mu,m}\left(\hat\mu_N\right)\right)\]
 and the term \ref{lign2} by 
\[2\sqrt{n}\E[\|\dd_{\mu,m}-\dd_{\hat\mu_{N},m}\|_{\infty,\x}].\]

This is proved in the three following lemmata.
\epv

\lm[Study of term \ref{lign2}]\label{Lemme u_E}
We have
\begin{align*}
&W_1\left(\LLL(\sqrt{n}W_1(\dd_{\hat\mu_{N},m}(\hat\mu_n),\dd_{\hat\mu'_{N},m}(\hat\mu'_n))),\LLL(\sqrt{n}W_1(\dd_{\mu,m}(\hat\mu_n),\dd_{\mu,m}(\hat\mu'_n)))\right)\leq\\
&2\sqrt{n}\E[\|\dd_{\mu,m}-\dd_{\hat\mu_{N},m}\|_{\infty,\x}].
\end{align*}
\elm
\pv
To bound this $L_1$-Wasserstein distance, we choose as a transport plan the law of the random vector 
\[(\sqrt{n}W_1(\dd_{\hat\mu_{N},m}(\hat\mu_n),\dd_{\hat\mu'_{N},m}(\hat\mu'_n)),\sqrt{n}W_1(\dd_{\mu,m}(\hat\mu_n),\dd_{\mu,m}(\hat\mu'_n))),\]
 with $\hat\mu_n$, $\hat\mu'_n$, $\hat\mu_{N-n}$ and $\hat\mu'_{N-n}$ independent empirical measures of law $\mu$. 
Then the $L_1$-Wasserstein distance is bounded by:
\[\E[|\sqrt{n}W_1(\dd_{\hat\mu_{N},m}(\hat\mu_n),\dd_{\hat\mu'_{N},m}(\hat\mu'_n))-\sqrt{n}W_1(\dd_{\mu,m}(\hat\mu_n),\dd_{\mu,m}(\hat\mu'_n))|],\]
which is not bigger than:
\[\sqrt{n}\E[W_1(\dd_{\hat\mu_{N},m}(\hat\mu_n),\dd_{\mu,m}(\hat\mu_n))+W_1(\dd_{\hat\mu'_{N},m}(\hat\mu'_n),\dd_{\mu,m}(\hat\mu'_n))].\]
We bound the term $\E[W_1(\dd_{\hat\mu_{N},m}(\hat\mu_n),\dd_{\mu,m}(\hat\mu_n))]$ by $\E[\|\dd_{\mu,m}-\dd_{\hat\mu_{N},m}\|_{\infty,\x}]$, thanks to Lemma \ref{lemme du stage}.
\epv

\lm[Study of term \ref{lign3}]\label{Lemme terme 3}
We have
\begin{align*}
&W_1\left(\LLL(\sqrt{n}W_1(\dd_{\mu,m}(\hat\mu_n),\dd_{\mu,m}(\hat\mu'_n)),\LLL(\sqrt{n}W_1(\dd_{\mu,m}(\mu^*_n),\dd_{\mu,m}(\mu'^*_n))|\hat\mu_N)\right)\leq\\
&2\sqrt{n}W_1(\dd_{\mu,m}(\mu),\dd_{\mu,m}(\hat\mu_N)).
\end{align*}
\elm
\pv
Let $\pi$ be the optimal transport plan associated to $W_1\left(\dd_{\mu,m}(\mu),\dd_{\mu,m}(\hat\mu_N)\right)$; see the definition of the $L_1$-Wasserstein with transport plans.

From a $n$-sample of law $\pi$, we get two empirical distributions $\dd_{\mu,m}(\hat\mu_n)$ and $\dd_{\mu,m}(\mu^*_n)$. Independently, from another $n$-sample of law $\pi$, we get $\dd_{\mu,m}(\hat\mu'_n)$ and $\dd_{\mu,m}(\mu'^*_n)$. 

The $L_1$-Wasserstein distance is then bounded by: 
\[\sqrt{n}\E_{\pi^{\otimes n}\otimes\pi^{\otimes n}}[W_1(\dd_{\mu,m}(\hat\mu_n),\dd_{\mu,m}(\mu^*_n))+W_1(\dd_{\mu,m}(\hat\mu'_n),\dd_{\mu,m}(\mu'^*_n))].\]
Now remark that, if we denote $\hat\mu_n=\sum_{i=1}^n\frac1n\delta_{Y_i}$ and $\mu^*_n=\sum_{i=1}^n\frac1n\delta_{Z_i}$, we have:
\begin{align*}
W_1(\dd_{\mu,m}(\hat\mu_n),\dd_{\mu,m}(\mu^*_n))&=\int_{t=0}^{+\infty}\left|\frac{1}{n}\sum_{i=1}^n\1_{\dd_{\mu,m}(Y_i)\leq t}-\frac{1}{n}\sum_{i=1}^n\1_{\dd_{\mu,m}(Z_i)\leq t}\right|\dd t\\
&\leq\frac{1}{n}\sum_{i=1}^n\int_{t=0}^{+\infty}\left|\1_{\dd_{\mu,m}(Y_i)\leq t}-\1_{\dd_{\mu,m}(Z_i)\leq t}\right|\dd t\\
&=\frac{1}{n}\sum_{i=1}^n\left|\dd_{\mu,m}(Y_i)-\dd_{\mu,m}(Z_i)\right|.
\end{align*}
So, the $L_1$-Wasserstein distance is not bigger than 
\[2\sqrt{n}\E[|\dd_{\mu,m}(Y)-\dd_{\mu,m}(Z)|],\]
with $(\dd_{\mu,m}(Y),\dd_{\mu,m}(Z))$ of law $\pi$,
so we get the upper bound:
\[2\sqrt{n}\left(W_1(\dd_{\mu,m}(\mu),\dd_{\mu,m}(\hat\mu_N)\right).\]
\epv

\lm[Study of term \ref{lign4}]
We have
\begin{align*}
&W_1\left(\LLL(\sqrt{n}W_1(\dd_{\mu,m}(\mu^*_n),\dd_{\mu,m}(\mu'^*_n))|\hat\mu_N),\LLL(\sqrt{n}W_1(\dd_{\hat\mu_N,m}(\mu^*_n),\dd_{\hat\mu_N,m}(\mu'^*_n))|\hat\mu_N)\right)\leq\\
&2\sqrt{n}\|\dd_{\mu,m}-\dd_{\hat\mu_N,m}\|_{\infty,\x}.
\end{align*}
\elm
\pv
It is the same proof as for the first lemma, except that $\hat\mu_N$ is fixed.
\epv

\lm\label{lemme du stage}
Let $\nu$, $\mu$ and $\mu'$ be some measures over some metric space $(\x,\delta)$, we have:
\[W_1(\dd_{\mu,m}(\nu),\dd_{\mu',m}(\nu))\leq \int_\x |\dd_{\mu,m}(x)-\dd_{\mu',m}(x)| \dd\nu(x) \leq\|\dd_{\mu,m}-\dd_{\mu',m}\|_{\infty,\Supp(\nu)}.\]
\elm
\pv
We chose the transport plan $(\dd_{\mu,m}(Y),\dd_{\mu',m}(Y))$ for $Y$ of law $\nu$.
\epv

Thanks to Proposition \ref{Stab DTM} and to the fact that the distance to a measure is 1-Lipschitz, we can derive another upper bound depending only on the $L_1$-Wasserstein distance between the measure $\mu$ and its empirical versions:

\cor
\label{Other bound for validity}
The quantity $W_1\left(\LLL_{N,n,m}(\mu,\mu),\LLL^*_{N,n,m}(\hat\mu_N,\hat\mu_N)\right)$ is upper bounded by:
\[2\frac{\sqrt{n}}{m}\E\left[W_1\left(\hat\mu_{N},\mu\right)\right]+2\sqrt{n}\left(1+\frac{1}{m}\right) W_1(\hat\mu_N,\mu).\]
\ecor

The rates of convergence of the $L_1$-Wasserstein distance between a Borel probability measure on the Euclidean space $\R^d$ and its empirical version are faster when the dimension $d$ is low; see \cite{Fournier}. Thus, we prefer to use the first bound for regular measures. In this case, we use rates of convergence for the distance to a measure, derived in \cite{Chazal}. For regular measures, in some cases, the bound in Lemma \ref{validity} is better than the bound in Corollary \ref{Other bound for validity}.

\subsection{An asymptotic result with the convergence to the law of $\|\G_{\mu,m}-\G'_{\mu,m}\|_1$}
\label{Conv loi pont brownien}
\label{C2}

\paragraph*{Proof of Lemma \ref{Asymptotic law to bridge}:}
The random function $\sqrt{n}\left(F_{\dd_{\mu,m}(\mu)}-F_{\dd_{\mu,m}(\hat\mu_n)}\right)$ converges weakly in $L_1$ to some gaussian prossess $\G_{\mu,m}$ with covariance kernel $\kappa(s,t)=F_{\dd_{\mu,m}(\mu)}(s)\left(1-F_{\dd_{\mu,m}(\mu)}(t)\right)$ for $s\leq t$; see \cite{delBarrio} or part 3.3 of \cite{Bobkov}. Thanks to Theorem 2.8 in \cite{Billingsley}, since $L_1\times L_1$ is separable and $\hat\mu_n$ and $\hat\mu'_n$ are independent, the random vector
\[\left(\sqrt{n}\left(F_{\dd_{\mu,m}(\mu)}-F_{\dd_{\mu,m}(\hat\mu_n)}\right),\sqrt{n}\left(F_{\dd_{\mu,m}(\mu)}-F_{\dd_{\mu,m}(\hat\mu'_n)}\right)\right)\]
converges weakly to $(\G_{\mu,m},\G'_{\mu,m})$ with $\G_{\mu,m}$ and $\G'_{\mu,m}$ independent Gaussian processes. Since the map $(x,y)\mapsto x-y$ is continuous in $L_1$, the mapping theorem states that $\sqrt{n}\left(F_{\dd_{\mu,m}(\hat\mu'_n)}-F_{\dd_{\mu,m}(\hat\mu_n)}\right)$ converges weakly to the Gaussian process $\G_{\mu,m}-\G'_{\mu,m}$ in $L_1$. Once more we use the mapping theorem with the continuous map $x\mapsto\|x\|_1$ and the definition of the $L_1$-Wasserstein distance as the $L_1$-norm of the cumulative distribution functions to get that:
\[\sqrt{n}W_1(\dd_{\mu,m}(\hat\mu_n),\dd_{\mu,m}(\hat\mu'_n))\leadsto \|\G_{\mu,m}-\G'_{\mu,m}\|_1.\]

We then get the convergence of moments following the same method as for Theorem 2.4 in \cite{delBarrio}.
We have the bound $\E[\|t\mapsto\1_{\dd_{\mu,m}(X_i)\leq t}-\1_{\dd_{\mu,m}(Y_i)\leq t}\|_1]\leq\Diam_\mu<\infty$. Moreover, the random function $\sqrt{n}\left(F_{\dd_{\mu,m}(\hat\mu'_n)}-F_{\dd_{\mu,m}(\hat\mu_n)}\right)$ converges weakly to the gaussian process $\G_{\mu,m}-\G'_{\mu,m}$ in $L_1$. So, thanks to Theorem 5.1 in \cite{deAcosta} (cited in \cite{Araujo} p.136), we have:
\[\E[\sqrt{n}W_1(\dd_{\mu,m}(\hat\mu_n),\dd_{\mu,m}(\hat\mu'_n))]\rightarrow\E[\|\G_{\mu,m}-\G'_{\mu,m}\|_1].\]

We deduce that:
\[W_1\left(\LLL\left(\sqrt{n}W_1\left(\dd_{\mu,m}\left(\hat\mu_n\right),\dd_{\mu,m}\left(\hat\mu'_n\right)\right)\right),\LLL\left(\|\G_{\mu,m}-\G'_{\mu,m}\|_1\right)\right)\rightarrow 0.\]

Moreover, we have the bound:
\[W_1\left(\LLL\left(\sqrt{n}W_1\left(\dd_{\mu,m}\left(\hat\mu_n\right),\dd_{\mu,m}\left(\hat\mu'_n\right)\right)\right),\LLL_{N,n,m}(\mu,\mu)\right)\leq2\sqrt{n}\E[\|\dd_{\mu,m}-\dd_{\hat\mu_{N},m}\|_{\infty,\x}].\]

So, if $\sqrt{n}\E\left[\|\dd_{\mu,m}-\dd_{\hat\mu_{N},m}\|_{\infty,\x}\right]\rightarrow 0$ when $N\rightarrow\infty$, we have that:
\[W_1\left(\LLL_{N,n,m}(\mu,\mu),\LLL\left(\|\G_{\mu,m}-\G'_{\mu,m}\|_1\right)\right)\rightarrow 0.\]
Finally, with the same arguments as for Lemma \ref{validity}, we get that:
\begin{align*}
&W_1\left(\LLL^*_{N,n,m}(\hat\mu_N,\hat\mu_N),\LLL\left(\|\G_{\mu,m}-\G'_{\mu,m}\|_1\right)\right)\leq\\
&W_1\left(\LLL\left(\sqrt{n}W_1\left(\dd_{\mu,m}\left(\hat\mu_n\right),\dd_{\mu,m}\left(\hat\mu'_n\right)\right)\right),\LLL\left(\|\G_{\mu,m}-\G'_{\mu,m}\|_1\right)\right)\\
&+2\sqrt{n}W_1\left(\dd_{\mu,m}(\mu),\dd_{\mu,m}\left(\hat\mu_N\right)\right)+2\sqrt{n}\|\dd_{\mu,m}-\dd_{\hat\mu_N,m}\|_{\infty,\x}.
\end{align*}
$\blacksquare$

\paragraph*{Proof of Proposition 17:}
Let $\epsilon<\alpha$ and $\eta$ be two positive numbers.

The probability $\p_{(\mu,\nu)}\left(\phi_N=1\right)$ is upper bounded by 
\[\p\left(\sqrt{n}W_1\left(\dd_{\hat\mu_{N},m}(\hat\mu_n),\dd_{\hat\nu_{N},m}(\hat\nu_n)\right)\geq\qu_{\alpha+\epsilon}-\eta\right)+\p\left(\hat\qu_\alpha<\qu_{\alpha+\epsilon}-\eta\right).\]
With a drawing, we see that $\p\left(\hat\qu_\alpha<\qu_{\alpha+\epsilon}-\eta\right)$ is upper bounded by \[\p\left(W_1\left(\LLL\left(\frac12\|\G_{\mu,m}-\G'_{\mu,m}\|_1+\frac12\|\G_{\nu,m}-\G'_{\nu,m}\|_1\right),\LLL^*\right)\geq\epsilon\eta\right),\]
where $\LLL^*=\frac12\LLL^*_{N,n,m}(\hat\mu_N,\hat\mu_N)+\frac12\LLL^*_{N,n,m}(\hat\nu_N,\hat\nu_N)$.

Thanks to the weak convergences in Lemma \ref{Asymptotic law to bridge} of the paper and the Portmanteau lemma, $\limsup_{N\rightarrow\infty}\p_{(\mu,\nu)}\left(\phi_N=1\right)$ is thus upper bounded by 
\[\p\left(\frac12\|\G_{\mu,m}-\G'_{\mu,m}\|_1+\frac12\|\G_{\nu,m}-\G'_{\nu,m}\|_1\geq\qu_{\alpha+\epsilon}-\eta\right).\]

We now make $\eta$ and $\epsilon$ go to zero and under the continuity assumption, $\limsup_{N\rightarrow\infty}\p_{(\mu,\nu)}\left(\phi_N=1\right)\leq\alpha$.

As well, we get that $\liminf_{N\rightarrow\infty}\p_{(\mu,\nu)}\left(\phi_N=1\right)\geq\alpha$.

\subsection{The case of measures supported on a compact subset of $\R^d$}
\label{section compact supported}

\paragraph*{Proof of part 2 of Proposition \ref{prop compact}:}

We may assume that the diameter $\Diam_\mu$ of the support of the measure $\mu$ equals 1. Indeed, if we apply a dilatation to the measure to make the diameter of its support be equal to 1, then the quantity $W_1\left(\LLL_{N,n,m}(\mu,\mu),\LLL^*_{N,n,m}(\hat\mu_N,\hat\mu_N)\right)$ is simply multiplied by the parameter of the dilatation. By using Corollary \ref{Other bound for validity} and Theorem 1 of \cite{Fournier}, we have a bound for the expectation:
$$
\E\left[W_1\left(\LLL_{N,n,m}(\mu,\mu),\LLL^*_{N,n,m}(\hat\mu_N,\hat\mu_N)\right)\right]\leq
\begin{cases}
C\frac{\sqrt{n}}{m}N^{-\frac1d}&\text{ if d>2}\\
C\frac{\sqrt{n}}{m}N^{-\frac12}\log(1+N)&\text{ if d=2}\\
C\frac{\sqrt{n}}{m}N^{-\frac12}&\text{ if d<2}\\
\end{cases}
$$
for some positive constant $C$ depending on $\mu$.
$\blacksquare$

\paragraph*{Proof of part 3 of Proposition \ref{prop compact}:}

First remark that for $\lambda>1$, 
\[\p\left(W_1\left(\LLL_{N,n,m}(\mu,\mu),\LLL^*_{N,n,m}(\hat\mu_N,\hat\mu_N)\right)\geq\lambda\right)=0\]
under the assumption $\Diam_\mu=1$. We thus focus on values of $\lambda$ not bigger than 1.
In this case, with the Theorem 2 of \cite{Fournier}, we get easily that:

\begin{align*}
\p&\left(W_1\left(\LLL_{N,n,m}(\mu,\mu),\LLL^*_{N,n,m}(\hat\mu_N,\hat\mu_N)\right)\geq\lambda\right)\leq\\
&\begin{cases}
C\exp\left(-C'\left(\lambda\frac{N^\frac1dm}{\sqrt{n}}-C''\right)^d\right)&\text{ for d>2}\\
C\exp\left(-C'\left(\frac{\frac{\sqrt{N}m}{\sqrt{n}}\lambda-C''\sqrt{\frac{N}{N-n}}\log(1+N-n)}{\log\left(2+\frac{2\sqrt{N}}{\frac{\sqrt{N}m}{\sqrt{n}}\lambda-C''\sqrt{\frac{N}{N-n}}\log(1+N-n)}\right)}\right)^2\right)&\text{ for d=2}\\
C\exp\left(-C'\left(\lambda\frac{\sqrt{N}m}{\sqrt{n}}-C''\right)^2\right)&\text{ for d<2}\\
\end{cases}
\end{align*}

for some positive constants $C$, $C'$ and $C''$ depending on $\mu$.

We conclude the proof with the Borel--Cantelli lemma.
$\blacksquare$

\paragraph*{Proof of part 1 of Proposition \ref{prop compact}:}

We need to show that under the assumption $\rho>\frac{\max\{d,2\}}{2}$, the following properties are satisfied:
\[\sqrt{n}\E[\|\dd_{\mu,m}-\dd_{\hat\mu_{N},m}\|_{\infty,\x}]\rightarrow 0,\]
\[\sqrt{n}W_1(\dd_{\mu,m}(\mu),\dd_{\mu,m}(\hat\mu_N))\rightarrow 0\text{ a.e.},\]
and
\[\sqrt{n}\|\dd_{\mu,m}-\dd_{\hat\mu_N,m}\|_{\infty,\x}\rightarrow 0\text{ a.e.}.\]

We treat the case $d>2$. The cases $d<2$ and $d=2$ are similar.

Thanks to Theorem 1 of \cite{Fournier}, there is some positive constant $C$ depending on $\mu$ such that for $N$ big enough: \[\E[W_1(\hat\mu_{N},\mu)]\leq C{N}^{-\frac{1}{d}}.\]
Thus, thanks to part 2 of Proposition \ref{prop compact}, the quantity $\sqrt{n}\E[\|\dd_{\mu,m}-\dd_{\hat\mu_{N},m}\|_{\infty,\x}]$ goes to zero if $\frac{\sqrt{n}}{m}N^{-\frac{1}{d}}$ goes to zero when $N$ goes to infinity. So, this convergence occurs under the assumption $\rho>\frac{d}{2}$.

We get from Theorem 2 of \cite{Fournier} that for $x\leq 1$, there are some positive constants $C$ and $c$ depending on $\mu$ such that:
\[\p(W_1(\hat\mu_N,\mu)\geq x)\leq C\exp(-cNx^d).\]
We use this inequality with $x=\frac{m}{\sqrt{n}}\frac{1}{K}$ for positive integers $K$. Thanks to the Borel--Cantelli lemma, under the assumption $\rho>\frac{d}{2}$, we get that:
\[\frac{\sqrt{n}}{m}W_1(\mu,\hat\mu_N)\rightarrow0\text{ a.e.}.\]
So, thanks to Proposition \ref{Stab DTM}, the third property is true.

To finish, remark that $\dd_{\mu,m}(\hat\mu_N)$ is the empirical measure associated to $\dd_{\mu,m}(\mu)$. Once more we use Theorem 2 of \cite{Fournier} and get that for $x\leq1$, $\p(\sqrt{n}W_1(\dd_{\mu,m}(\hat\mu_N),\dd_{\mu,m}(\mu))\geq x)\leq C\exp(-c\frac{N}{n}x^2)$. Thanks to the Borel--Cantelli lemma, under the assumption $\rho>1$, the a.e. convergence to zero of $\sqrt{n}W_1\left(\dd_{\mu,m}(\mu),\dd_{\mu,m}\left(\hat\mu_N\right)\right)$ occurs.
$\blacksquare$

\subsection{The case of $(a,b)$-standard measures}
\label{C3}

Let $\mu$ be a Borel probability measure supported on a connected compact subset $\x$ of $\R^d$. We assume this measure to be $(a,b)$-standard for some positive numbers $a$ and $b$. In this part, we derive rates of convergence in probability and in expectation for the quantity $\|\dd_{\hat\mu_N,m}-\dd_{\mu,m}\|_{\infty,\x}$. Thanks to these results, we can derive upper bounds and rates of convergence in expectation for $W_1\left(\LLL_{N,n,m}(\mu,\mu),\LLL^*_{N,n,m}(\hat\mu_N,\hat\mu_N)\right)$. We finally propose a choice for the parameter $N$ depending on $n$ for which the weak convergences $\LLL_{N,n,m}(\mu,\mu)\leadsto\|\G_{\mu,m}-\G'_{\mu,m}\|_1$ and $\LLL^*_{N,n,m}(\hat\mu_N,\hat\mu'_N)\leadsto\|\G_{\mu,m}-\G'_{\mu,m}\|_1$ occur.

\subsubsection{Upper bounds for $\p(\sqrt{n}\|\dd_{\hat\mu_N,m}-\dd_{\mu,m}\|_{\infty,\x}\geq\lambda)$}

We use the bounds given in Theorem 1 of \cite{Chazal}, with the bound for the modulus of continuity given by \textbf{Lemma 3} in \cite{Chazal}: $\omega(h)=\left(\frac ha\right)^{\frac1b}$. We directly get the following lemma:

\lm[Upper bound for $|\dd_{\hat\mu_N,m}(x)-\dd_{\mu,m}(x)|$]
Let $x$ be a fixed point in $\x$ and $\lambda$ a positive number. We have,
\begin{align*}
&\frac12\p(|\dd_{\hat\mu_N,m}(x)-\dd_{\mu,m}(x)|\geq\lambda)\leq\\
&\exp\left(-2a^\frac2bNm^{\frac{2b-2}{b}}\lambda^2\right)+\exp\left(-\frac{a}{2^{b-1}}N^{\frac{b+1}{2}}m^b\lambda^b\right)+\exp\left(-a^\frac1bN^{\frac{b+1}{2b}}m\lambda\right).
\end{align*}
\elm

In order to derive an upper bound for $\|\dd_{\hat\mu_N,m}-\dd_{\mu,m}\|_{\infty,\x}$, like in \cite{Chazal}, we use the fact that the function distance to a measure is 1-Lipschitz and that $\x$ is compact, which means that we can compute a bound by upper-bounding the difference $|\dd_{\hat\mu_N,m}(x)-\dd_{\mu,m}(x)|$ over a finite number of points $x$ of $\x$. Thanks to the following lemma, the minimal number of points needed for this purpose is not bigger than $\frac{(4\Diam_\mu\sqrt d+\lambda)^d}{\lambda^d}$:

\lm
\label{Nhausdorff}
Let $\mu$ is a measure supported on $\x$ a compact subset of $\R^d$, and for $\lambda>0$ denote $N\left(\mu,\lambda\right)=\inf\{N\in\N,\ \exists\ x_1,x_2\ldots x_N\in\x,\ \bigcup_{i\in[\![1,N]\!]}\B(x_i,\lambda)\supset\x\}$.
Then, we have:
\[N\left(\mu,\lambda\right)\leq\frac{\left(\Diam_\mu\sqrt{d}+\lambda\right)^d}{\lambda^d}.\]
\elm
\pv
The idea is to put a grid on the hypercube containing $\x$ with edges of length $\Diam_\mu$. The grid is a union of small hypercubes with edges of length equal to $\frac{\lambda}{\sqrt{d}}$, so that the number of such small hypercubes into which the big one is split is not superior to $\left(\frac{\Diam_\mu\sqrt d}{\lambda}+1\right)^d$. 

Then, we decide that each time the intersection between $\x$ and some small hypercube is non-empty, we keep one of the elements of the intersection. We denote $x_i$ the element associated to the $i$-th hypercube. Finally, each point $x$ in $\x$ belongs to a small hypercube, and its distance to the corresponding $x_i$ is smaller than $\sqrt{\sum_{k=1}^d\frac{\lambda^2}{d}}=\lambda$.
\epv

We thus derive upper bounds for $\sqrt{n}\|\dd_{\hat\mu_N,m}-\dd_{\mu,m}\|_{\infty,\x}$:

\prop[Upper bound for $\sqrt{n}\|\dd_{\hat\mu_N,m}-\dd_{\mu,m}\|_{\infty,\x}$]
\label{Upper bound norm infty}
We have,
\begin{align*}
&\frac{\lambda^d}{2\left(4\Diam_\mu\sqrt{d}+\lambda\right)^d}\ \p(\sqrt{n}\|\dd_{\hat\mu_N,m}-\dd_{\mu,m}\|_{\infty,\x}\geq\lambda)\leq\\
&\exp\left(-\frac{a^\frac2b}{2}\frac{Nm^{\frac{2b-2}{b}}}{n}\lambda^2\right)+\exp\left(-\frac{a}{2^{2b-1}}\frac{N^{\frac{b+1}{2}}m^b}{n^\frac{b}{2}}\lambda^b\right)+\exp\left(-\frac{a^\frac1b}{2}\frac{N^{\frac{b+1}{2b}}m}{n^\frac12}\lambda\right).
\end{align*}
\eprop
\pv
Since the function distance to a measure is 1-Lipschitz, we get that:
\[\|\dd_{\hat\mu_N,m}-\dd_{\mu,m}\|_{\infty,\x}\leq\frac{\lambda}{2}+\sup_i\{|\dd_{\hat\mu_N,m}(x_i)-\dd_{\mu,m}(x_i)|\},\]
for the family $(x_i)_i$ associated to a grid which sides are of length equal to $\frac{\lambda}{4\sqrt{d}}$.
We can thus bound the probability $\p(\|\dd_{\hat\mu_N,m}-\dd_{\mu,m}\|_{\infty,\x}\geq\lambda)$ by:
\[\sum_{i=1}^{N\left(\mu,\frac{\lambda}{4}\right)}\p\left(|\dd_{\hat\mu_N,m}(x_i)-\dd_{\mu,m}(x_i)|\geq\frac{\lambda}{2}\right),\]
with $N\left(\mu,\frac{\lambda}{4}\right)\leq\frac{\left(4\Diam_\mu\sqrt{d}+\lambda\right)^d}{\lambda^d}$ thanks to Lemma \ref{Nhausdorff}.
\epv
\subsubsection{Upper bounds for the expectation $\E[\|\dd_{\hat\mu_N,m}-\dd_{\mu,m}\|_{\infty,\x}]$}

In order to get upper bounds for $\E[\|\dd_{\hat\mu_N,m}-\dd_{\mu,m}\|_{\infty,\x}]$, we use the same trick as used in \cite{Chazal}, which is:

\lm
Let $X$ a random variable such that:
\[\p(X\geq\lambda)\leq1\wedge D\lambda^{-q}\exp(-c\lambda^s)\]
for some integers $q$ and $s$ and some $D>0$.

We have:
\[\E[X]\leq\left(\frac{\ln c}{c}\right)^{\frac1s}\left(\frac qs\right)^{\frac1s}\left[1+D\left(\frac qs\right)^{\frac{-q-s}{s}}\frac{(\ln c)^{\frac{-q-s}{s}}}s\right].\]

More particularly, if $c\geq \exp{D^{\frac{s}{q+s}}\frac{s}{q}}$, then:
\[\E[X]\leq2\left(\frac{\ln c}{c}\right)^{\frac1s}\left(\frac qs\right)^{\frac1s}.\]
\elm

\pv
For any $\lambda_0>0$, that we can choose as $\lambda_0=\frac{[\ln K]^\frac1s}{c^{\frac1s}}$, we get that:
\begin{align*}
&\E[X]\leq \lambda_0 + \int_{\lambda_0}^\infty D\lambda^{-q}\exp(-c\lambda^s)\dd\lambda\\
& \leq \lambda_0+D\frac{\lambda_0^{-q-s+1}}{cs}\exp{-c\lambda_0^s}\\
& = \frac{[\ln K]^\frac1s}{c^{\frac1s}} + D \frac{[\ln K]^{\frac{-q-s+1}{s}}}{sc c^{\frac{-q-s+1}{s}}} \frac1K\\
& = \frac{[\ln K]^\frac1s}{c^{\frac1s}} + \frac{[\ln K]^{\frac 1s}}{c^\frac1s} D \frac{[\ln K]^{\frac{-q-s}{s}}}{s c^{\frac{-q}{s}}} \frac1K\\
& = \frac{[\ln K]^\frac1s}{c^{\frac1s}} \left[1+D\frac{[\ln K]^{\frac {-q-s}s}}{s K c^{\frac{-q}{s}}}\right]
\end{align*}
Finally, if we choose $ K=c^{\frac qs}$, we get:
\[\E[X]\leq\left(\frac qs\right)^{\frac 1s}\left[\frac{\ln c}{c}\right]^{\frac 1s}\left[1+D\left[\frac{q}{s}\right]^{\frac{-q-s}{s}}\frac{(\ln c)^{\frac{-q-s}{s}}}{s}\right].\]
\epv

From this lemma, we can derive the following lemma.

\lm
\label{Majoration esperance}
We have,
\begin{align*}
&\E[\sqrt{n}\|\dd_{\hat\mu_N,m}-\dd_{\mu,m}\|_{\infty,\x}]\leq\\
& \Box'_1 \frac{n^\frac12}{N^\frac12m^{\frac{b-1}{b}}}\left(\log\left(\frac{Nm^{\frac{2b-2}{b}}}{n}\right)\right)^{\frac12}+\\
& \Box'_2 \frac{n^\frac12}{N^{\frac{b+1}{2b}}m}\left(\log\left(\frac{N^{\frac{b+1}{2}}m^b}{n^\frac{b}{2}}\right)\right)^\frac1b+\\
& \Box'_3 \frac{n^\frac12}{N^{\frac{b+1}{2b}}m}\log\left(\frac{N^{\frac{b+1}{2b}}m}{n^\frac12}\right).
\end{align*}
for some constants $\Box$ depending on $a$ and $b$.
\elm

\subsubsection{Upper bounds for the expectation of $W_1\left(\LLL_{N,n,m}(\mu,\mu),\LLL^*_{N,n,m}(\hat\mu_N,\hat\mu_N)\right)$}

\paragraph*{{Proof of part 2 of Proposition \ref{prop std}:}}

For all $\lambda>0$, for any measure $\mu$ Ahlfors $b$-regular with parameters $(a,\infty)$ supported on a connected compact subset of $\R^d$, we can use Lemma \ref{validity} and Lemma \ref{Majoration esperance} together with the rates of convergence of the $L_1$-Wasserstein distance between empirical and true distribution in \cite{Bobkov} to get the following result.

If $m\geq\frac12$, then for $n$ big enough we have, for some constants $\Box$ depending on $a$ and $b$:
\begin{align*}
\E&\left[W_1\left(\LLL_{N,n,m}(\mu,\mu),\LLL^*_{N,n,m}(\hat\mu_N,\hat\mu_N)\right)\right]\leq\\
& \Box'_1 \frac{n^\frac12}{{(N)}^\frac12m^{\frac{b-1}{b}}}\left(\log\left(\frac{Nm^{\frac{2b-2}{b}}}{n}\right)\right)^{\frac12}\\
&+ \Box'_2 \frac{n^\frac12}{{(N)}^{\frac{b+1}{2b}}m}\left(\log\left(\frac{N^{\frac{b+1}{2}}m^b}{n^\frac{b}{2}}\right)\right)^\frac1b\\
&+ \Box'_3 \frac{n^\frac12}{{(N)}^{\frac{b+1}{2b}}m}\log\left(\frac{N^{\frac{b+1}{2b}}m}{n^\frac12}\right)\\
&+ \Box'_4 \frac{n^\frac12}{N^\frac12}.
\end{align*}

\subsubsection{Convergence to the law of $\|\G_{\mu,m}-\G'_{\mu,m}\|_1$}

\paragraph*{Proof of part 1 of Proposition \ref{prop std}:}
In order to get these two results, we use Lemma \ref{Asymptotic law to bridge}. The convergence to zero of $\sqrt{n}\E[\|\dd_{\mu,m}-\dd_{\hat\mu_{N-n},m}\|_{\infty,\x}]$ is a direct consequence of Lemma \ref{Majoration esperance}. We can derive a bound of its rate of convergence in $n^{\frac12-\frac{\rho}{2}}$, up to a logarithm term. The a.e. convergence of $\sqrt{n}W_1(\dd_{\mu,m}(\mu),\dd_{\mu,m}(\hat\mu_N))$ to zero is derived as in the proof of Proposition \ref{prop compact}, with the assumption $\rho>1$. Finally, the a.e. convergence of $\sqrt{n}\|\dd_{\mu,m}-\dd_{\hat\mu_N,m}\|_{\infty,\x}$ to zero is a consequence of Proposition \ref{Upper bound norm infty} and of the Borel--Cantelli lemma. It occurs under the assumption $\rho>1$.
$\blacksquare$

\subsection{The power of the test}
\label{C4}

\paragraph*{Proof of Proposition 21}

\lm
\label{quantiles tries}
Let $\alpha$, $\kappa$ be two positive numbers and $\LLL$ and $\LLL^*$ two laws of real random variables. We denote $\qu_\alpha$ (respectively $\qu_\alpha^*$) the $\alpha$-quantile of the law $\LLL$ (respectively $\LLL^*$).
If $W_1(\LLL,\LLL^*)<\kappa$ then:
\[\qu_\alpha^*\leq2\frac{\kappa}{\alpha}+\qu_{\frac{\alpha}{2}}.\] 
\elm
\pv
With a drawing, since the $L_1$-norm between $F_{\LLL}$ and $F_{\LLL^*}$ is smaller than $\kappa$, we have:
\[F_{\LLL^*}\left(\qu_{\frac{\alpha}{2}}+2\frac{\kappa}{\alpha}\right)>1-\alpha.\]
\epv
In this part we assume that $m$ is fixed in $[0,1]$ and $N=cn^\rho$ for some $\rho>1$ and $c>0$.

Recall that our aim is to upper bound the type II error, that is:
\[\p_{(\mu,\nu)}\left(\sqrt{n}W_1\left(\dd_{\hat\mu_{N},m}(\hat\mu_n),\dd_{\hat\nu_{N},m}(\hat\nu_n)\right)<\hat\qu_\alpha\right).\]

For some $\kappa=n^\gamma$ with $\gamma$ in $\left[0,\frac12\right)$ to be chosen later, we first upper bound the quantile $\hat\qu_\alpha$ with high probability.

As noticed in the proof of Lemma \ref{Asymptotic law to bridge}, the law of $\sqrt{n}W_1\left(\dd_{\mu,m}(\hat\mu_n),\dd_{\mu,m}(\hat\mu'_n)\right)$ converges to $\LLL(\|\G_{\mu,m}-\G'_{\mu,m}\|_1)$, there is also the convergence of the first moments. So, for $n$ big enough, we have: 
\[W_1(\LLL\left(\sqrt{n}W_1\left(\dd_{\mu,m}(\hat\mu_n),\dd_{\mu,m}(\hat\mu'_n)\right)\right),\LLL(\|\G_{\mu,m}-\G'_{\mu,m}\|_1))\leq1.\]
Then, under the assumption 
 \[W_1(\LLL\left(\sqrt{n}W_1\left(\dd_{\mu,m}(\hat\mu_n),\dd_{\mu,m}(\hat\mu'_n)\right)\right),\LLL^*_{N,n,m}(\hat\mu_N,\hat\mu_N))\leq\kappa,\]
 we have 
 \[W_1(\LLL(\|\G_{\mu,m}-\G'_{\mu,m}\|_1),\LLL^*_{N,n,m}(\hat\mu_N,\hat\mu_N))\leq\kappa+1.\]
 We can do the same thing for $\nu$.
 Thus we get that for $n$ big enough and under the previous assumptions:
 \[W_1\left(\frac12\LLL(\|\G_{\mu,m}-\G'_{\mu,m}\|_1)+\frac12\LLL(\|\G_{\nu,m}-\G'_{\nu,m}\|_1),\frac12\LLL^*_{N,n,m}(\hat\mu_N,\hat\mu_N)+\frac12\LLL^*_{N,n,m}(\hat\nu_N,\hat\nu_N)\right)\leq\kappa+1.\]
 And thanks to Lemma \ref{quantiles tries}, \[\hat\qu_\alpha\leq\tilde\qu_{\frac{\alpha}{2}}+2\frac{\kappa+1}{\alpha},\]
with $\tilde\qu_\alpha$ the $\alpha$-quantile of the law $\frac12\LLL(\|\G_{\mu,m}-\G'_{\mu,m}\|_1)+\frac12\LLL(\|\G_{\nu,m}-\G'_{\nu,m}\|_1)$.

 We need to remark that with similar arguments as for Lemma \ref{validity}, we have:
 \begin{align*}
 &W_1\left(\LLL\left(\sqrt{n}W_1\left(\dd_{\mu,m}(\hat\mu_n),\dd_{\mu,m}(\hat\mu'_n)\right)\right),\LLL^*_{N,n,m}(\hat\mu_N,\hat\mu_N)\right)\leq\\
 &2\sqrt{n}\Diam_\mu\|F_{\dd_{\mu,m}(\mu)}-F_{\dd_{\mu,m}(\hat\mu_N)}\|_{\infty,(0,\Diam_\mu)}+2\frac{\sqrt{n}}{m}W_1(\mu,\hat\mu_N).
 \end{align*}\\
 
Now remark that
\begin{align*}
&\sqrt{n}W_1\left(\dd_{\hat\mu_{N},m}(\hat\mu_n),\dd_{\hat\nu_{N},m}(\hat\nu_n)\right)\geq\sqrt{n}W_1\left(\dd_{\mu,m}(\mu),\dd_{\nu,m}(\nu)\right)\\
&-\sqrt{n}W_1\left(\dd_{\hat\mu_{N},m}(\hat\mu_n),\dd_{\mu,m}(\mu)\right)-\sqrt{n}W_1\left(\dd_{\hat\nu_{N},m}(\hat\nu_n),\dd_{\nu,m}(\nu)\right),
\end{align*}

but as well, thanks to Lemma \ref{lemme du stage}, the definition of the $L_1$-Wasserstein distance as the $L_1$-norm between the cumulative distribution functions and to Proposition \ref{Stab DTM}:
\begin{align*}
&\sqrt{n}W_1\left(\dd_{\hat\mu_{N},m}(\hat\mu_n),\dd_{\mu,m}(\mu)\right)\leq\\
&\frac{\sqrt{n}}{m}W_1(\mu,\hat\mu_{N})+\sqrt{n}\Diam_{\mu,m}\|F_{\dd_{\mu,m}(\hat\mu_n)}-F_{\dd_{\mu,m}(\mu)}\|_{\infty,(0,\Diam_\mu)},
\end{align*}
with $\Diam_{\mu,m}$ the diameter of the support of the measure $\dd_{\mu,m}(\mu)$.
So, we can finally upper bound $\p_{(\mu,\nu)}\left(\sqrt{n}W_1\left(\dd_{\hat\mu_{N-n},m}(\hat\mu_n),\dd_{\hat\nu_{N-n},m}(\hat\nu_n)\right)<\hat\qu_\alpha\right)$ by 
\begin{align*}
&\p\left(\sqrt{n}\Diam_\mu\|F_{\dd_{\mu,m}(\mu)}-F_{\dd_{\mu,m}(\hat\mu_N)}\|_{\infty,(0,\Diam_\mu)}\geq\frac{\kappa}{4}\right)+\\
&\p\left(\sqrt{n}\Diam_\nu\|F_{\dd_{\nu,m}(\nu)}-F_{\dd_{\nu,m}(\hat\nu_N)}\|_{\infty,(0,\Diam_\nu)}\geq\frac{\kappa}{4}\right)+\\
&2\p\left(\frac{\sqrt{n}}{m}W_1(\mu,\hat\mu_{N})\geq\frac{\kappa}{4}\right)+2\p\left(\frac{\sqrt{n}}{m}W_1(\nu,\hat\nu_{N})\geq\frac{\kappa}{4}\right)+\\
&\p\left(\|F_{\dd_{\mu,m}(\hat\mu_n)}-F_{\dd_{\mu,m}(\mu)}\|_{\infty,(0,\Diam_\mu)}\geq\frac{W_1(\dd_{\mu,m}(\mu),\dd_{\nu,m}(\nu))}{2\Diam_{\mu,m}}-\frac{\tilde\qu_{\frac{\alpha}{2}}}{2\Diam_{\mu,m}\sqrt{n}}-\frac{(4+\alpha)\kappa+4}{4\Diam_{\mu,m}\alpha\sqrt{n}}\right)+\\
&\p\left(\|F_{\dd_{\nu,m}(\hat\nu_n)}-F_{\dd_{\nu,m}(\nu)}\|_{\infty,(0,\Diam_\nu)}\geq\frac{W_1(\dd_{\mu,m}(\mu),\dd_{\nu,m}(\nu))}{2\Diam_{\nu,m}}-\frac{\tilde\qu_{\frac{\alpha}{2}}}{2\Diam_{\nu,m}\sqrt{n}}-\frac{(4+\alpha)\kappa+4}{4\Diam_{\nu,m}\alpha\sqrt{n}}\right).
\end{align*}

For all positive $\epsilon$, for $n$ big enough, remark that the sum of the last two terms can be bounded thanks to the DKW-Massart inequality \cite{Massart}, by 
\[4\exp\left(-\frac{W_1^2\left(\dd_{\mu,m}(\mu),\dd_{\nu,m}(\nu)\right)}{(2+\epsilon)\max\left\{\Diam_\mu^2,\Diam_\nu^2\right\}}n\right).\] 

Remark also that thanks to the DKW-Massart inequality, the first term can be upper bounded by 
\[2\exp\left(-\frac{1}{8\Diam_\mu^2}cn^{\rho-1+2\gamma}\right).\]
The second term is similar. Thanks to Theorem 2 in \cite{Fournier}, the third term is upper bounded by 
\[c_1\exp\left(-c_2m^dn^{\rho+d\gamma-\frac{d}{2}}\right),\]
for some fixed constants $c_1$ and $c_2$. The remaining terms are similar.

Since $\rho>1$, we can choose a positive $\gamma$ satisfying: $\gamma<\frac12$, $\rho+d\gamma-\frac{d}{2}>1$ and $\rho-1+2\gamma>1$. So the two last expressions are negligible in comparison to the first one.

So, for $n$ big enough, $\p_{(\mu,\nu)}\left(\sqrt{n}W_1\left(\dd_{\hat\mu_{N},m}(\hat\mu_n),\dd_{\hat\nu_{N},m}(\hat\nu_n)\right)<\hat\qu_\alpha\right)$ is upper bounded by \[4\exp\left(-\frac{W_1^2\left(\dd_{\mu,m}(\mu),\dd_{\nu,m}(\nu)\right)}{3\max\left\{\Diam_{\mu,m}^2,\Diam_{\nu,m}^2\right\}}n\right).\]
$\blacksquare$

\subsection{Numerical illustrations}

In this section, we give details on the simulations presented in Section \ref{illustration}. Recall that we consider the measure $\mu_{v}$, that is, the distribution of the random vector $(R\sin(v R)+0.03 N,R\cos(v R)+0.03 N')$ with $R$, $N$ and $N'$ independent random variables; $N$ and $N'$ from the standard normal distribution and $R$ uniform on $(0,1)$.\\

From the measure $\mu_{10}$ we get a $N$-sample $P=\{X_1,X_2,\ldots,X_{N}\}$, where $N=2000$. As well, we get a $N$-sample $Q=\{Y_1,Y_2,\ldots,Y_{N}\}$ from the measure $\mu_{20}$. It leads to the empirical measures $\hat\mu_{10,N}$ and $\hat\mu_{20,N}$. On Figure \ref{signaturem005}, we plot the cumulative distribution function of the measure $\dd_{\hat\mu_{10,N},m}(\hat\mu_{10,N})$, that is, the function $F$ defined for all $t$ in $\R$ by the proportion of the  $X_i$ in $P$ satisfying $\dd_{\hat\mu_{10,N},m}(X_i)\leq t$. It approximates the true cumulative distribution function associated to the DTM-signature $\dd_{\mu,m}(\mu)$. As well, we plot the cumulative distribution function of the measure $\dd_{\hat\mu_{20,N},m}(\hat\mu_{20,N})$.
Observe that the signatures are different. Thus, for the choice of parameter $m=0.05$, the DTM-signature discriminates well between the measures $\mu_{10}$ and $\mu_{20}$.\\

In Figure \ref{Bootstrapm005}, for $m=0.05$ and $n=20$, we first generate $N_{MC}=1000$ independent realisations of the random variable $\sqrt{n}W_1(\dd_{\hat\mu_{10,N-n},m}(\hat\mu_{10,n}),\dd_{{\hat{\mu}'}_{10,N-n},m}({\hat{\mu}'}_{20,n}))$, where $\hat{\mu}_{10,N}$ and ${\hat{\mu}'}_{10,N}$ are independent empirical measures from $\mu_{10}$, $\hat\mu_{10,N}=\frac{n}{N}\hat\mu_{10,n}+\frac{N-n}{N}\hat\mu_{10,N}$ and ${\hat{\mu}'}_{10,N}=\frac{n}{N}{\hat{\mu}'}_{10,n}+\frac{N}{N}{\hat{\mu}'}_{10,N-n}$. We plot the empirical cumulative distribution function associated to this $N$-sample. As well, from two fixed $N$-samples from the law $\mu_{10}$, $P$ and $Q$, we generate a set $boot$ of $N_{MC}$ random variables, as explained in the Algorithm in Section \ref{algo}, and we plot its cumulative distribution function. Remark that the too cumulative distribution functions are close. It means that the $\alpha$-quantile of the distribution of the test statistic is well approximated by the $\alpha$-quantile of the bootstrap distribution.\\

The Figure \ref{table} is obtained by applying the test \textbf{DTM} and the test \textbf{KS} to two independent $N$-samples, 1000 times independently, and by averaging the number of rejections of the hypothesis $H_0$. For the type-I error, the $N$-samples are both from $\mu_{10}$, as for the power, a sample is from $\mu_{10}$ and the other one from $\mu_v$.
\end{appendix}
\end{document}